\documentclass[preprint,amsmath,amssymb,superscriptaddress,longbibliography,aps,floatfix,footinbib]{revtex4-2}

\usepackage{graphicx}
\usepackage{tabularx}
\usepackage[usenames,dvipsnames]{color}
\usepackage{soul}
\usepackage{bm}
\usepackage{amsmath}
\usepackage{lineno}
\usepackage{gensymb}

\usepackage{times}
\usepackage{amsfonts}
\usepackage{mathrsfs}
\usepackage{graphicx}
\usepackage{dcolumn}
\usepackage{bm}
\usepackage{color}

\usepackage[colorlinks,bookmarks=false,citecolor=blue,linkcolor=red,urlcolor=blue]{hyperref}
\usepackage{multirow}
\usepackage{physics}

\begin{document}

\title{Anomalous spin-optical helical effect in Ti-based kagome metal}

\author{Federico Mazzola**}
\email{federico.mazzola@spin.cnr.it}
\affiliation{CNR-SPIN, c/o Complesso di Monte S. Angelo, IT-80126 Napoli, Italy}

\author{Wojciech Brzezicki**}
\email{w.brzezicki@uj.edu.pl}
\affiliation{Institute of Theoretical Physics, Jagiellonian University, ulic, S. \L{}ojasiewicza 11, PL-30348 Krak\'ow, Poland}
\affiliation{International Research Centre MagTop, Institute of Physics, Polish Academy of Sciences, Aleja Lotnik\'ow 32/46, PL-02668 Warsaw, Poland}

\author{Chiara Bigi}
\affiliation{Synchrotron SOLEIL, L’Orme des Merisiers, D\'epartementale 128, F-91190 Saint-Aubin, France}

\author{Armando Consiglio}
\affiliation{Istituto Officina dei Materiali, Consiglio Nazionale delle Ricerche, Trieste I-34149, Italy}

\author{Luciano Jacopo D'Onofrio}
\affiliation{CNR-SPIN, c/o Universit\'a di Salerno, IT-84084 Fisciano (SA), Italy}

\author{Maria Teresa Mercaldo}
\affiliation{Dipartimento di Fisica ``E. R. Caianiello", Universit\`a di Salerno, IT-84084 Fisciano (SA), Italy}

\author{Adam Kłosiński}
\affiliation{Institute of Theoretical Physics, Faculty of Physics, University of Warsaw, Pasteura 5, PL-02093 Warsaw, Poland}

\author{Fran\c cois Bertran}
\affiliation{Synchrotron SOLEIL, L’Orme des Merisiers, D\'epartementale 128, F-91190 Saint-Aubin, France}

\author{Patrick Le F\`evre}
\affiliation{Univ Rennes, IPR Institut de Physique de Rennes, UMR 6251, F-35000 Rennes, France}

\author{Oliver J. Clark}
\affiliation{School of Physics and Astronomy, Monash University, Clayton, Victoria 3800, Australia}

\author{Mark T. Edmonds}
\affiliation{School of Physics and Astronomy, Monash University, Clayton, Victoria 3800, Australia}

\author{Manuel Tuniz}
\affiliation{Dipartimento di Fisica, Universita degli studi di Trieste, 34127, Trieste, Italy}

\author{Alessandro De Vita}
\affiliation{Fritz Haber Institut der Max Planck Gesellshaft, Faradayweg 4--6, 14195 Berlin, Germany\looseness=-1}

\author{Vincent Polewczyk}
\affiliation{Universit\'e Paris-Saclay, UVSQ, CNRS, GEMaC, 78000, Versailles, France}

\author{Jeppe B. Jacobsen}
\affiliation{Nanoscience Center, Niels Bohr Institute, University of Copenhagen, 2100 Copenhagen, Denmark}

\author{Henrik Jacobsen} 
\affiliation{European Spallation Source ERIC - Data Management and Software Center, 2800 Kgs.~Lyngby, Denmark}

\author{Jill A. Miwa}
\affiliation{Department of Physics and Astronomy, Interdisciplinary Nanoscience Center, Aarhus University, 8000 Aarhus C, Denmark}

\author{Justin W. Wells}
\affiliation{Department of Physics and Centre for Materials Science and Nanotechnology, University of Oslo (UiO), 0318 Oslo, Norway.}

\author{Anupam Jana}
\affiliation{CNR-IOM Istituto Officina dei Materiali, I-34139 Trieste, Italy}

\author{Ivana Vobornik}
\affiliation{CNR-IOM Istituto Officina dei Materiali, I-34139 Trieste, Italy}

\author{Jun Fujii}
\affiliation{CNR-IOM Istituto Officina dei Materiali, I-34139 Trieste, Italy}

\author{Niccolo Mignani}
\affiliation{Dipartimento di Fisica, Politecnico di Milano, Piazza Leonardo Da Vinci 32, Milano 20133, Italy}

\author{Narges Samani Tarakameh}
\affiliation{Dipartimento di Fisica, Politecnico di Milano, Piazza Leonardo Da Vinci 32, Milano 20133, Italy}

\author{Alberto Crepaldi}
\affiliation{Dipartimento di Fisica, Politecnico di Milano, Piazza Leonardo Da Vinci 32, Milano 20133, Italy}

\author{Giorgio Sangiovanni}
\affiliation{Institute for Theoretical Physics and Astrophysics, University of W\"urzburg, D-97074 W\"urzburg, Germany}

\author{Anshu Kataria}
\affiliation{Dipartimento di Scienze Matematiche, Fisiche e Informatiche, Universitá di Parma, I-43124 Parma, Italy}

\author{Tommaso Morresi}
\affiliation{European Centre for Theoretical Studies in Nuclear Physics and Related Areas (ECT*), Fondazione Bruno Kessler, Trento, Italy}

\author{Samuele Sanna}
\affiliation{Department of Physics and Astronomy, University of Bologna, 40127 Bologna, Italy}

\author{Pietro Bonf\'a}
\affiliation{Dipartimento di Scienze Matematiche, Fisiche e Informatiche, Universitá di Parma, I-43124 Parma, Italy}

\author{Brenden R. Ortiz}
\affiliation{Materials Department, University of California Santa Barbara, Santa Barbara, California 93106, USA}

\author{Ganesh Pokharel}
\affiliation{Materials Department, University of California Santa Barbara, Santa Barbara, California 93106, USA}

\author{Stephen D. Wilson}
\affiliation{Materials Department, University of California Santa Barbara, Santa Barbara, California 93106, USA}

\author{Domenico Di Sante}
\affiliation{Department of Physics and Astronomy, University of Bologna, 40127 Bologna, Italy}

\author{Carmine Ortix}
\affiliation{Dipartimento di Fisica ``E. R. Caianiello", Universit\`a di Salerno, IT-84084 Fisciano (SA), Italy}

\author{Mario Cuoco}\email{mario.cuoco@spin.cnr.it}
\affiliation{CNR-SPIN, c/o Universit\'a di Salerno, IT-84084 Fisciano (SA), Italy}

\begin{abstract}



\bf{The kagome lattice stands as a rich platform for hosting a wide array of correlated quantum phenomena, ranging from charge density waves and superconductivity to electron nematicity and loop current states. Direct detection of loop currents in kagome systems has remained a formidable challenge due to their intricate spatial arrangements and the weak magnetic field signatures they produce. This has left their existence and underlying mechanisms a topic of intense debate. In this work, we uncover a hallmark reconcilable with loop currents: spin handedness-selective signals that surpass conventional dichroic, spin, and spin-dichroic responses. We observe this phenomenon in the kagome metal CsTi$_3$Bi$_5$ and we call it the anomalous spin-optical helical effect. This effect arises from the coupling of light’s helicity with spin-orbital electron correlations, providing a groundbreaking method to visualize loop currents in quantum materials. Our discovery not only enriches the debate surrounding loop currents but also paves the way for new strategies to exploit the electronic phases of quantum materials via light-matter interaction.}
\end{abstract}

\maketitle
\noindent

Materials with kagome lattices provide an exceptional platform for exploring emergent correlated quantum phenomena, including charge density waves, superconductivity, and electronic nematicity \cite{Yi2023, Jiang2024, Nie2022, Hu2023, Schwemmer2024, Neupert2022, Kang2022, kang2023}. Kagome metals exhibit distinct electronic structure features, such as itinerant Dirac-like states and flat bands, with established topologically non-trivial properties \cite{Kang2020, Disante2023, Tuniz2023, Li2021, Liu2020}. The ‘135’ kagome family (AB$_3$C$_5$, with A: Rb, Cs, K; B: Ti, V; and C: Bi, Sb) has been extensively studied due to its rich variety of electronic correlated phases \cite{Hu_2022, Zhao_2021, Zheng2022}, alongside the intriguing potential for time-reversal symmetry breaking induced by loop currents \cite{Le2024, Guo2024, Guo2022, Tazai2023, Christensen2022}.

Loop current states, known for their violation of time-reversal symmetry, are marked by electrons circulation along closed lattice paths, generating a magnetic flux. This flux, while intrinsic to the nature of these states, is difficult to detect due to its low amplitude and complex spatial distribution \cite{Bounoua2020, Bounoua2022, Tang2018}. Despite extensive theoretical and experimental exploration, loop currents in kagome systems remain contentious regarding their origin and nature \cite{Tazai2023, Christensen2022, Dong2023, Liege2024}. These states are hypothesized to arise from bond-order fluctuations \cite{Tazai2023} or symmetry reduction induced by charge density waves, evidenced by observations of chiral transport switchable with magnetic fields \cite{Guo2022}. Recent neutron diffraction experiments by Liège \textit{et al.} have detected signatures of loop currents in CsV$_3$Sb$_5$, suggesting electron circulation confined to vanadium sites \cite{Liege2024}. However, weak signals raise questions about the authenticity of the observed time-reversal symmetry breaking, underscoring the need for further investigation into the nature of these loop currents. The latter can exhibit identifiable signatures linked to the symmetry of spin and orbital patterns within the current flow but direct experimental confirmation has remained challenging.

When loop currents exhibit strong spin-orbital correlations, they can induce an optical phenomenon involving the interaction between light helicity and electron spin. This effect, which we term spin-orbital helical coupling, generates signals surpassing those of conventional spin, dichroic, spin, and spin-dichroic responses, presenting a novel and promising method for detecting elusive quantum phases. In this work, motivated by a theory of spin-orbital correlated loop current phases, we employ single-handed polarized spin- and angle-resolved photoelectron spectroscopy (spin-ARPES) to reveal the existence of spin-optical helical coupling in CsTi$_3$Bi$_5$. Our findings provide robust signatures which are compatible with time-reversal symmetry breaking induced by loop currents marked by nonstandard spin-orbital quadrupole correlations.

\begin{figure}
\includegraphics[width=1.0\columnwidth]{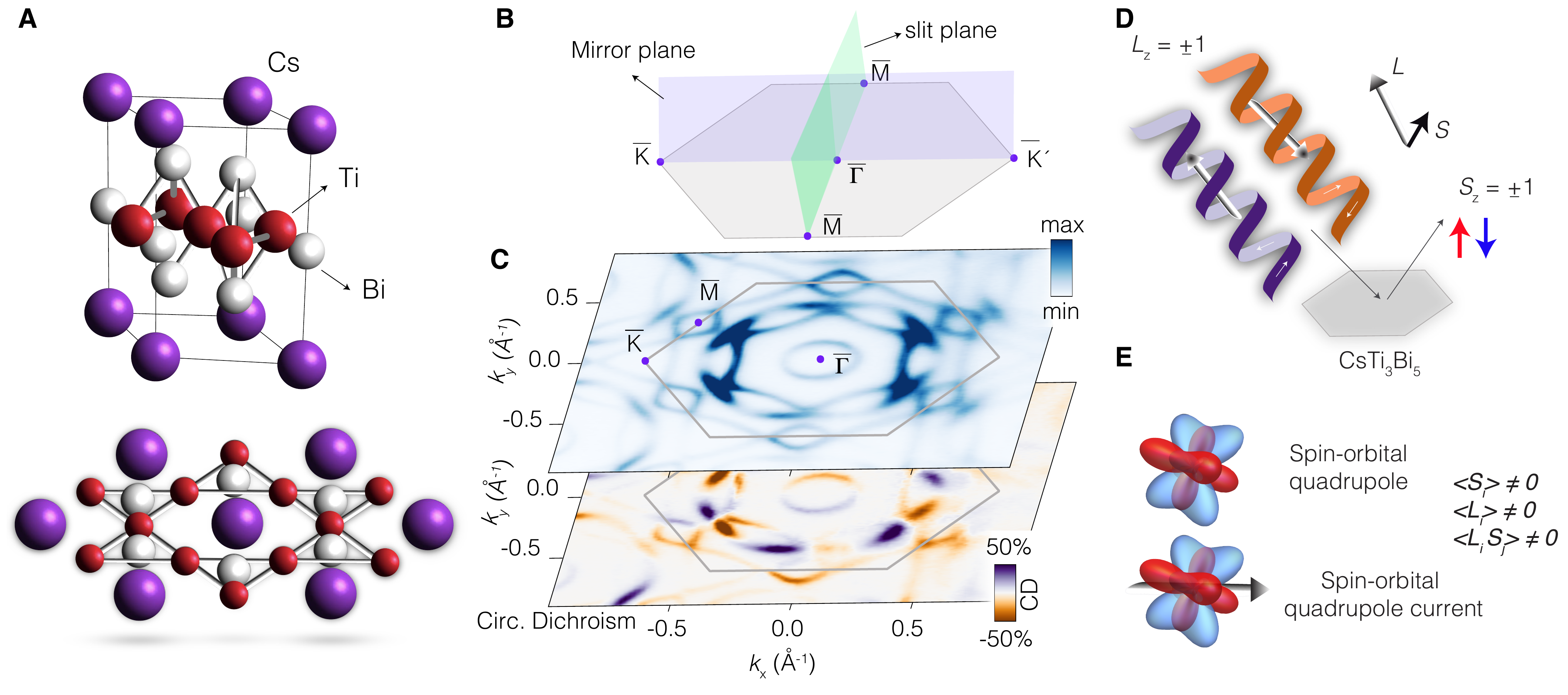}
  \caption{\textbf{a.} Side view and top view crystal structure of CsTi$_3$Bi$_5$ showing the typical kagome geometry featured by the Ti atoms (red). \textbf{b.} Experimental geometry: the incoming radiation comes within the mirror plane of the crystal and the measurements are collected orthogonal to it, along the plane of the analyzer's slit. \textbf{c.} Fermi surface map of CsTi$_3$Bi$_5$ obtained by summing up circular right and left polarization (monochromatic color map) and subtracting them (purple-orange color map). In the experimental geometry used, the circular dichroism signal is well-defined. \textbf{d.} Schematic of the spin-resolved circularly polarized ARPES: the helicity of the light can couple to the orbital angular momentum and the spin of electrons. By using a spin-detector, we can select the spin of the photoemitted electrons. \textbf{e.} Example of spin-orbital quadrupoles and of currents generated by them. The breaking of the relevant spin and orbital quantities is also depicted.}
  \label{fig:1}
\end{figure}

Single crystals of CsTi$_3$Bi$_5$ were grown using a flux-based growth technique, as detailed in reference \cite{growth} and the methods section. The crystal structure, depicted in Fig.\ref{fig:1}a, is characteristic of the '135' kagome family, featuring Ti atoms (red) that form a frustrated kagome network, Bi atoms (white) residing within the same plane, and alkali Cs (purple) positioned above and below this plane. While CsTi$_3$Bi$_5$ shares several features with its V-based sister compounds, it does not undergo a charge density wave transition, nor was superconductivity observed in our samples, contrary to what reported in a recent study \cite{Yang_2024}. This characteristic renders CsTi$_3$Bi$_5$ a simpler system for studying intrinsic properties, where correlations and frustrations may act as primary driving forces.

The samples were cleaved under ultrahigh vacuum conditions ($1 \times 10^{-10}$ mbar), and all photoemission experiments were performed with synchrotron light directed at the sample's surfaces within one of the crystal's mirror plane (purple plane in Fig.\ref{fig:1}b). The analyser's slit was oriented orthogonally to this plane (green shades in Fig.\ref{fig:1}b). This experimental configuration ensures that the geometrical contributions to the circular dichroism are well-defined, yielding anti-symmetric outcomes in a perfectly mirror and time-reversal symmetric scenario.

\begin{figure}
\includegraphics[width=1.0\columnwidth]{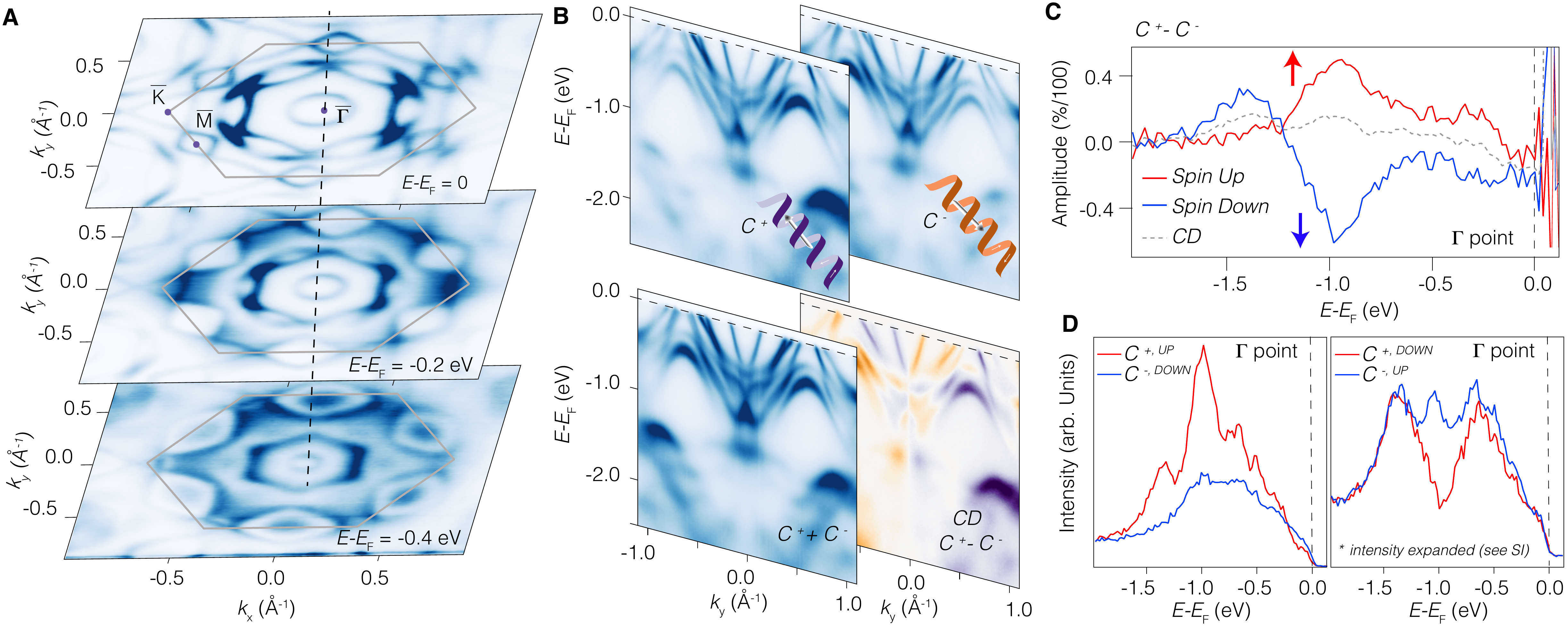}
  \caption{\textbf{a.} Constant energy maps obtained by summing both circular right and left polarization. The kagome signatures are visible in reciprocal space, similar to other members of the '135' family. \textbf{b.} Energy versus momenta maps showing the typical electronic structure of CsTi$_3$Bi$_5$ obtained with both circular right and left polarization (up), and by summing and subtracting them (down). \textbf{c.} Spin-resolved CP-ARPES (up spin in red, down spin in blue) collected from the $\Gamma$ point, together with spin-integrated signal (gray). \textbf{d.} Single handed polarization signals showing the breaking of time reversal symmetry.}
  \label{fig:2}
\end{figure}

Fermi surfaces collected with circularly polarized angle-resolved photoelectron spectroscopy (CP-ARPES), at photon energy where the spectral features were prominent and the circular dichroism from ARPES more symmetric (See also supplementary information figures 1-7 for data collected with various photon energies and polarizations, including $k_z$ dispersions and constant energy contours). The measurements reveal the characteristic kagome geometry, featuring a circular pocket centered at $\bar{\Gamma}$, two hexagonal sheets rotated by 30 degrees relative to each other, triangular features around $\bar{K}$, and diamond-shaped pockets centered at $\bar{M}$ (see Fig.\ref{fig:1}c and Fig.\ref{fig:2}a for constant energy maps) \cite{Yang2023, Wang2023, Liu2023, Jiang2023}. The observed circular dichroism of the Fermi surface, collected without spin-detector, indicates that the system conserves overall time-reversal and mirror symmetries (see Fig.\ref{fig:1}c). Along the $\bar{\Gamma}$-$\bar{K}$ high-symmetry direction, the circular dichroism is weak and negligible within the experimental resolution, specifically regarding the degree of asymmetry in light polarization (approximately 10$\%$, see also Ref.\cite{Disante2023}). This condition, with net negligible circular dichroism at specific $k$-points, is a reasonable choice when performing spin-resolved circular dichroism measurements (See Fig.\ref{fig:2}d), because could minimize possible scattering contributions and Daimon effect \cite{Boban_2024}, allowing us to visualize better spin-orbital contributions (Fig.\ref{fig:2}e). In addition, we anticipate that this experiment is performed exclusively for signals from the Brillouin zone center.

Recent studies have demonstrated that quantum systems with negligible circular dichroism can still exhibit non-vanishing contributions in spin-dichroic amplitudes, which can be linked to chirality and loop currents \cite{Mazzola2024, Fittipaldi2021}. For CsTi$_3$Bi$_5$, we have investigated such spin-dichroic amplitudes, exploiting spin-resolved CP-ARPES. To ensure comparability with prior literature \cite{Disante2023}, 
we used a controlled experimental geometry with measurements from the $\Gamma$ point using the spin-detectors (See methods). At first glance, the system behaves similarly to previously reported findings for CoSn \cite{X} and the '166' kagome family XV$_6$Sn$_6$ (X: Ho, Sc, Tb) \cite{Disante2023}: the spin-integrated CP-ARPES from $\Gamma$, as anticipated, shows small deviations from zero. The spin-resolved CP-ARPES from the same point, with up (red curve in Fig.\ref{fig:2}c) and down (blue curve in Fig.\ref{fig:2}c) spins, exhibits overall opposite signs. This result indicates that, within the experimental uncertainty of the polarization of the incoming radiation, time-reversal symmetry shows no significant anomalies. In earlier investigations, the sign-reversal of the spin-resolved CP-ARPES was enforced by mirror and time-reversal symmetry. This is reflected in the total dichroic response ($CD=C^+-C^-$), where $CD^{ \uparrow}$ transitions into $-CD^{\downarrow}$ upon application of the time-reversal operator $(TR)$. This symmetry relation was also observed in single-handed polarized measurements: $TR(C^{+, \uparrow})=C^{-,\downarrow}$ and $TR(C^{+,\downarrow})=C^{-,\uparrow}$. However, in the absence of symmetry constraints, the spin-dichroic left and right amplitudes may be different. The breaking of such a symmetry constraint for single handed light measurements is the way spin-optical helical coupling manifests.

The latter effect is indeed observed in CsTi$_3$Bi$_5$, exhibiting a remarkably large and time-reversal symmetry broken response to single-handed polarized light, as shown by the curves in Fig.\ref{fig:2}d. Such a significant signal indicates that the mirror symmetric character of electrons is broken and that substantial spin-orbital correlations are present, thereby supporting the compatibility with loop currents of spin-orbital nature, as will be demonstrated later. Importantly, as we show in supplementary figures 9 and 10, this result is independent on the choice of background and normalization used. We stress that while the spin-optical coupling with the single helicity of light is consistently realized, the large asymmetry observed at $\Gamma$ serves as compelling evidence for strong spin-orbital correlations that break time-reversal symmetry. Theoretically, one can expect to observe such a symmetry breaking also in the spin signals, even if with amplitudes smaller than the spin-orbital ones. Whilst the latter appear strong in our experiments and are evident along the full energy range collected for the energy distribution curves from $\Gamma$, the bare spin-signal (circularly unpolarized data) was less clear, not conclusive and difficult to disentangle from other spurious effects (See supplementary figure 11 and discussion). 

We also consider whether the observed spin-optical anomaly is accompanied  by the occurrence of a sizable magnetic moment. This aspect has been explored by means of muon spin rotation and relaxation experiments that have been conducted on powdered samples of CsTi$_3$Bi$_5$ (see SI for details). The analysis of the spectra, along with first principles simulations to assess the muon localization in the crystal, indicates that no static magnetic fields larger than 0.25 mT can be present at the muon site. This would potentially correspond to a dipolar contribution from Ti atoms originating from a magnetic moment smaller than $2 \times 10^{-3}$~$\mu_{\text B}$.

\begin{figure}
\includegraphics[width=0.75\columnwidth]{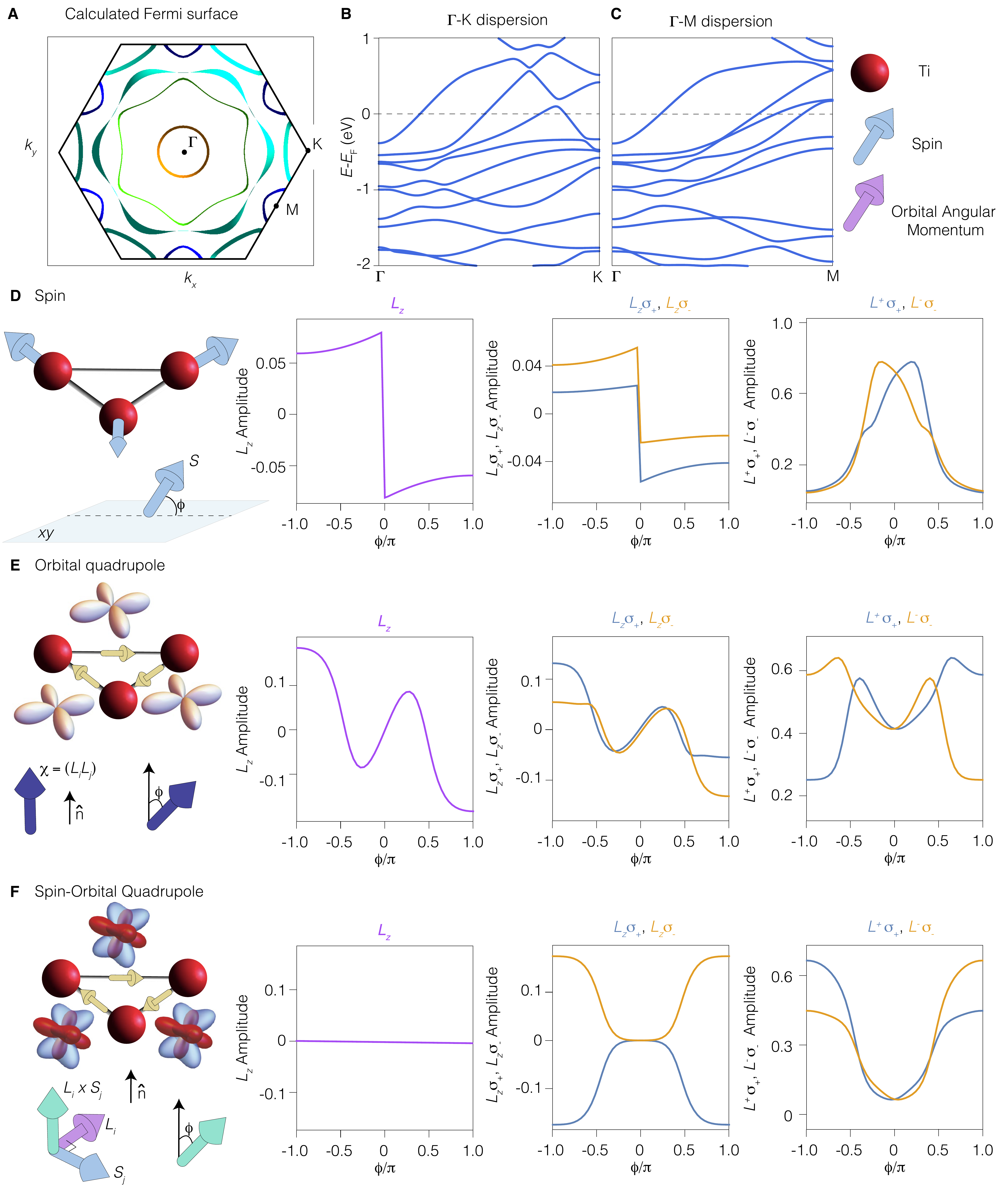}
\caption{\textbf{a.} Fermi surface of CsTi$_3$Bi$_5$ calculated via density functional theory. \textbf{b.} Electronic dispersion along the $\Gamma-K$ and $\Gamma-M$ directions of the Brillouin zone. \textbf{c.} Schematic of magnetic ordering with spin moments forming 120-degree angles in the plane; one spin orientation varies, introducing an out-of-plane component at angle $\phi$. Orbital polarization ($L_z$), spin-resolved projections ($L_z \sigma_{+}$, $L_z \sigma_{-}$), and left- ($L^+ \sigma_{-}$) and right-handed ($L^{-} \sigma_{+}$) amplitudes depend on $\phi$. \textbf{d.} Sketch of orbital quadrupole loop-current in the kagome unit cell, with three mirror-symmetric components parameterized by coupling $\chi$. The polar angle $\phi$ influences the strength of $L_x L_y$ relative to $L_x L_z$ and $L_y L_z$. $L_z$ is nonzero only at isolated $\phi$ points, with differing magnitudes for spin-resolved orbital polarizations. \textbf{e.} Illustration of spin-orbital quadrupole loop-current with $\bf{L} \times \bf{s}$ correlations. The out-of-plane amplitude varies with $\phi$, while planar components remain fixed. The response shows unique features: $L_z$ vanishes, and spin-up and spin-down dichroic components are equal but opposite, while left- and right-handed orbital polarizations differ in magnitude.}
\label{fig:3}
\end{figure}

We will now examine our experimental results from a theoretical perspective. We commence by analyzing the Fermi surface structure and the electronic dispersion near the Fermi level for the relevant bands investigated in the ARPES experiment. In Fig. \ref{fig:3}a, we present the Fermi surface obtained via density functional theory (DFT) (see Methods). The observed structure of the Fermi lines closely resembles the experimental observations (Fig. \ref{fig:1}c) regarding the number and character of the Fermi pockets around the high-symmetry positions in the Brillouin zone. The shape and size of these structures correlate with the distinct band dispersions along the $\Gamma-K$ (Fig. \ref{fig:3}b) and $\Gamma-M$ (Fig. \ref{fig:3}c) directions, respectively. According to our DFT study, the electronic states near the Fermi level predominantly exhibit character stemming from the five d-orbitals of titanium.

To analyze the dichroic and spin-dichroic responses, we have developed a two-dimensional tight-binding model of the electronic structure of CsTi$_3$Bi$_5$, focusing on the complete multiplet of the Ti d-orbitals ($d_{xy}$, $d_{yz}$, $d_{zx}$, $d_{x^{2}-y^{2}}$, $d_{3z^{2}-r^{2}}$). This allows us to effectively replicate the characteristics of the experimental Fermi surface. The electronic parameters were derived from Wannier projections of the DFT bands, utilizing the Ti d-orbitals as the basis (see Methods).

Considering both spin and orbital degrees of freedom alongside the tri-sublattice structure of the kagome lattice, we construct an effective model employing the orbital angular momentum $L=2$ for the five d-orbital configurations at each Ti site, along with a $s=\frac{1}{2}$ moment for the electron spin. Additionally, we introduce a pseudospin $T=1$ to distinguish the states on the three inequivalent atoms within the unit cell. The Ti-Ti hybridization processes within the unit cell and between neighboring unit cells, as well as the local crystalline field potential, can be represented as a tensor product of the operators associated with the $L$, $T$, and $s$ algebras (see Supplementary Information). We then incorporate various types of symmetry-breaking states to evaluate the dichroic, spin-dichroic, and the spin-resolved left- and right-handed circularly polarized amplitudes of the photoemission process, specifically focusing on values of the crystal wave vector corresponding to the center of the Brillouin zone. 
The main aim here is to compare the response of different types of symmetry-breaking states and to demonstrate that the observed experimental features can be uniquely associated with loop currents flowing within the unit cell and carrying a distinct type of spin-orbital quadrupolar pattern.
For each band eigenstate $|\psi_{n}\rangle$ evaluated at $\Gamma$, we determine the amplitude of the out-of-plane orbital moment ${L_z}=\langle \psi_{n}| \hat{L}_z |\psi_{n}\rangle$, the spin-projected orbital moment $L_z \sigma_{\pm}=\langle \psi_{n}| (1\pm \hat{s}_z) \hat{L}_z |\psi_{n}\rangle$ (i.e., the out-of-plane spin-up ($+$) and down ($-$) components as selected by the projector $(1\pm \hat{s}_z)$), and the spin-resolved left and right-handed orbitally polarized amplitudes expressed as ${L^{\pm} \sigma_{\pm}}=\langle \psi_{n}| (1\pm \hat{s}_z) \hat{L}_z (1\pm\hat{L}_z) |\psi_{n}\rangle$, respectively. For readability, we have omitted the band index $n$ in the definitions of these observables. In a manner analogous to the spin projector, the operator $\hat{L}_z (1 \pm \hat{L}_z)$ within the $(L=2)$ manifold selects orbital configurations with left-handed ($+$) or right-handed ($-$) characteristics, dictated by the associated intrinsic orbital angular momentum. These observables are directly related to the dichroic ($L_z$), spin-dichroic ($L_z \sigma_{\pm}$), and spin-resolved left- and right-handed (${L^{\pm} \sigma_{\pm}}$) amplitudes probed by ARPES, respectively.

Before delving deeper into the behavior of these physical quantities, it is noteworthy that for electronic configurations exhibiting time-reversal or mirror symmetry, the amplitudes $L_z$, $L_z \sigma_{\pm}$, and $L^{\pm} \sigma_{\pm}$ either vanish or are interconnected by symmetry conditions. This stems from the fact that the operators $\bf{\hat{L}}$ and $\bf{\hat{s}}$ change sign under time-reversal symmetry transformation and behave as pseudovectors under mirror transformation. Therefore, in the presence of time and mirror symmetries, the electronic states probed at $\Gamma$ are not anticipated to exhibit any distinctive anomalies in the dichroic, spin-dichroic, or spin-resolved amplitudes associated with left- and right-handed circularly polarized light.

Next, we consider various magnetic configurations that break both time and mirror symmetries. We start with a standard magnetic ordering in which the spin moments are arranged to create 120-degree angles in the plane, as illustrated schematically in Fig. \ref{fig:3}d. This magnetic configuration may arise due to the frustration inherent in the kagome lattice and results in the breaking of both time and mirror symmetries. We specifically allow one of the spin moments within the unit cell to exhibit an out-of-plane component, characterized by the angle $\phi$. This introduces a non-vanishing scalar spin chirality, $\eta$, defined as the expectation value of $\bf{s}_1 \cdot (\bf{s}_2 \times \bf{s}_3)$ (where labels 1, 2, and 3 correspond to the three titanium sites within the unit cell). When $\phi$ deviates from $0$, $\eta$ does not vanish, thereby breaking all mirror symmetries and any combinations of time and mirror symmetries. In this context, the variable $\phi$ is introduced to illustrate that the effects are not coincidental and do not arise solely in specific instances of broken symmetry states.

In Fig. \ref{fig:3}d, we investigate the behavior of the orbital polarization $L_z$ (left panel), the spin-resolved components $(L_z \sigma_{+}, L_z \sigma_{-})$ (middle panel), and the spin-resolved single-handedness polarizations $(L^{+} \sigma_{-}, L^{-} \sigma_{+})$ (right panel) for a representative eigenstate $|\psi_{n}\rangle$ at $\Gamma$. We find that $L_z$ is generally non-vanishing, except when considering states with spin moments lying entirely within the kagome plane. The amplitude of the spin-resolved orbital polarization typically exhibits unequal magnitudes between the spin-up and spin-down channels, $(L_z \sigma_{+}, L_z \sigma_{-})$, a behavior that is also consistent for the spin-resolved single-handedness polarizations. This qualitative behavior is characteristic of all electronic states at $\Gamma$.

We will now examine two representative states of loop currents that demonstrate mirror-broken orbital configurations (Fig. \ref{fig:3}e) and spin-orbital quadrupoles (Fig. \ref{fig:3}f). Given that spin and orbital angular momentum possess pseudovector characteristics with magnetic dipole properties, their combination can be classified as a spin-orbital quadrupole or an orbital quadrupole. Both spin and orbital angular momentum as well as the sublattice pseudospin are odd parity (i.e. they change sign) under time-reversal transformation. Consequently, currents carrying spin-orbital or orbital quadrupoles, which are even in time, violate time-reversal symmetry.

The loop-current state characterized by mirror-broken orbital quadrupoles within the unit cell can be represented by introducing the vector field $\hat{\xi}=(\hat{\xi}_x,\hat{\xi}_y,\hat{\xi}_z)$, where ${\hat{\xi}}_x=\hat{L}_y \hat{L}_z + \hat{L}_z \hat{L}_y$, with other components derived through index permutation. The symmetry-breaking orbital loop-current term within the unit cell can then be gnerally expressed as $\hat{J}_o=(\hat{T}_x+\hat{T}_y+\hat{T}_z)(\chi_x \hat{\xi}_x+\chi_y \hat{\xi}_y+\chi_z \hat{\xi}_z)$. 

Analogous to the magnetic state with aligned spin moments, we analyze the dichroic, spin-dichroic and single-handedness responses for the loop-current state exhibiting mirror-broken orbital quadrupoles, as shown in Fig. \ref{fig:3}e. For clarity, we focus on a representative configuration that elucidates the overall behavior of the spin optical response. This is accomplished by maintaining the amplitudes of the $(x,y)$ components, specifically setting $\chi_x = \chi_y = \chi_0$, while varying $\chi_z$ as $\chi_z = \chi_0 \cdot \cos(\phi)$. The parameter $\phi$ serves to modulate the relative strength of the $\chi_z$ term with respect to the $\chi_x$ and $\chi_y$ components. The overall trend indicates that the orbital polarization typically does not vanish (left panel Fig. \ref{fig:3}e). Furthermore, the spin-resolved responses lack any discernible symmetric behavior in amplitude or sign when comparing the spin-up and spin-down channels (middle panel Fig. \ref{fig:3}e) or the left and right orbitally polarized configurations (right panel Fig. \ref{fig:3}e).
Therefore, both the magnetic state and the orbital-quadrupole loop current phase display a response that contradicts the experimental findings, which show a nearly zero dichroic amplitude and a symmetric spin-dichroic response.

Next, we consider the case of a loop-current state featuring mirror- and rotation-broken spin-orbital quadrupole current (Fig. \ref{fig:3}f). The loop-current state, characterized by a mirror-broken spin-orbital quadrupole within the unit cell, can be introduced by considering the vector field obtained from the cross product of the spin and orbital angular momentum, $\bf{\hat{L}} \times \bf{\hat{s}}$. Just like the orbital loop current, various couplings can be associated with different types of spin-orbital quadrupolar distributions.  The symmetry breaking spin-orbital loop-current term, within the unit cell, can be then expressed as $\hat{J}_{so}=g_{so} (\hat{T}_x+\hat{T}_y+\hat{T}_z)[\bf{n} \cdot (\bf{\hat{L}} \times \bf{\hat{s}})]$ with ${\bf{n}}=(n_x,n_y,n_z)$ setting the director for the spin-orbital quadrupole distribution, and $g_{so}$ the strength of the loop-current state. The variation of the spin-orbital loop current is made through the angle $\phi$, i.e. the angle between the vector $\bf{n}$ and the cross product of $\bf{\hat{L}}$ and $\bf{\hat{s}}$.  This type of loop-current configuration exhibits a distinct response if compared to the spin aligned magnetic state or the orbital quadrupole loop-current: the orbital polarization $L_z$ is zero (left panel Fig. \ref{fig:3}f) for all values of $\phi$, the orbital moment for spin-up and spin-down configurations $(L_z \sigma_{+}, L_z \sigma_{-})$ have equal magnitudes but opposite signs (middle panel Fig. \ref{fig:3}f), and in contrast, the handedness and spin-resolved orbital configurations display a profile with amplitudes that are unrelated in size or sign at any angle $\phi$ that characterizes the spin-orbital loop current state (right panel Fig. \ref{fig:3}f). We would like to point out that this qualitative profile is representative of all the electronic states at $\Gamma$, although the distribution of amplitudes changes with energy. These distinctive features characterize the spin-orbital loop current states, which are not typically found in other electronic phases that break time or mirror symmetries. Notably, the behavior of these loop current states, despite violating both mirror and time-reversal symmetries, exhibits vanishing dichroism and symmetric amplitudes in the spin-dichroic channel, traits that are usually associated with systems that uphold time-reversal or mirror symmetries.
Hence, the dichroic, the spin-dichroic and the handedness responses for this loop current phase match very well with the ARPES experimental observations.

In conclusion, we have experimentally demonstrated the occurrence of an anomalous spin-optical helical effect in the '135' kagome system CsTi$_3$Bi$_5$, indicating the presence of a loop current phase marked by robust spin-orbital correlations. Our findings reveal this effect through signals from the $\Gamma$ point, directly attributable to spin-orbital quadrupoles circulating across the central hexagonal lattice of the kagome structure, challenging and advancing the prevailing theories regarding bond order in the literature \cite{Tazai2023}. Moreover, in the absence of charge density waves, the established loop currents are intrinsic features of the frustrated atomic network. 
In addition, the effect observed, in analogy to previous studies \cite{Mazzola2024}, could also appear within the first layers of the materials, making this approach a unique tool to explore this phenomenology. While a comprehensive investigation into the role of neighboring atoms in mediating spin asymmetry and the spin-optical helical anomaly extends beyond the current study, it opens promising avenues for future research. Such explorations could deepen our understanding of the intricate interplay between atomic arrangements, electronic structures, and spin-orbital correlations with potential implications for the exploitation of novel quantum phases in technologies based on spin-optical helical effects.

\bibliography{kagome}

\section{Methods}
High-purity single crystals of CsTi$_3$Bi$_5$ were synthesized using a conventional flux-based method, following the procedure detailed in \cite{growth}. The starting materials included cesium (liquid, Alfa 99.98 $\%$), titanium (powder, Alfa 99.9 $\%$), and bismuth (shot, Alfa 99.999 $\%$), which were combined in a stoichiometric ratio of 1:1:6. This mixture was placed in a tungsten carbide milling vial and subjected to one hour of milling in an argon atmosphere to produce the precursor powder. The powder was then transferred to an alumina crucible, which was sealed in a stainless-steel tube. The sealed tube was heated to 900$^\circ,$C for 10 hours, followed by slow cooling at a rate of 3$^\circ,$C per hour until reaching 500$^\circ,$C. After the process, the tube was opened in an argon-filled glove box, where the shiny, plate-like single crystals were carefully separated and stored under an inert gas atmosphere.\\

The circularly polarized spin-ARPES measurements were performed with light incident within one of the sample's mirror plane. This critical choice enables precise control over experimental conditions, allowing for effective probing of geometrical matrix elements. We collected energy-distribution curves from the $\Gamma$ point, where the geometrical matrix elements from the circular dichroism  are defined to be zero. Subsequently, we obtained energy versus momentum maps using a spin detector, akin to the ARPES results reported in Fig.\ref{fig:2}b for spin-integrated circular dichroism (see supplementary information figure 8). After aligning the crystals based on these maps, we extracted the energy distribution curve resolved in spin and circular polarization from the $\Gamma$ point, as illustrated in Fig.\ref{fig:2}c.

Electronic structure calculations were performed using the full-potential local-orbital (FPLO) code (v.21.00-61) \cite{PhysRevB.59.1743}. The unit cell has lattice constants of 5.82709 Å, 5.82709 Å, and 9.93612 Å. The exchange-correlation energy was parametrized within the local density approximation, following the Perdew-Wang 92 formulation \cite{PhysRevB.45.13244}. A $12\times12\times12$ $k$-grid was used to sample the Brillouin Zone, and the tetrahedron method was employed for integration. Calculations were performed in both the full-relativistic and non-relativistic frameworks.\\
A subsequent Wannier function model \cite{PhysRev.52.191, RevModPhys.34.645}, with symmetries accurately implemented, was constructed, considering projections onto the following states: Caesium $6s$; Titanium $3d_{z^2}$, $3d_{xz}$, $3d_{yz}$, $3d_{x^2 - y^2}$, $3d_{xy}$; and Bismuth $6p_z$, $6p_x$, $6p_y$, $6s$.

\noindent{\bf{\large Acknowledgements}}\\
F.M. greatly acknowledges the SoE action of PNRR, number SOE\_0000068 and and the funding by the European Union – NextGenerationEU, M4C2, within the PNRR project NFFA-DI, CUP B53C22004310006, IR0000015. M.C. and M.T.M. acknowledge support from the EU’s Horizon 2020 research and innovation program under Grant Agreement No. 964398 (SUPERGATE). M.C. and L.J.D'O. acknowledge support from PNRR MUR project PE0000023-NQSTI. M.T.M and C.O. acknowledge partial support by the Italian Ministry of Foreign Affairs and International Cooperation, Grants No. PGR12351 (UTLRAQMAT). W.B. acknowledges support by Narodowe Centrum Nauki (NCN, National Science Centre, Poland) Project No. 2019/34/E/ST3/00404 and partial support by the Foundation for Polish Science through the IRA Programme co-financed by EU within SG OP. A. C. acknowledges support from PNRR MUR project PE0000023-NQSTI. A.C., G.S. and D.D.S. acknowledge the Gauss Centre for Supercomputing e.V. (https://www.gauss-centre.eu) for funding this project by providing computing time on the GCS Supercomputer SuperMUC-NG at Leibniz Supercomputing Centre (https://www.lrz.de). The authors thank Prof. Riccardo Comin for the useful insights and discussions. We acknowledge SOLEIL for provision of synchrotron radiation facilities under proposal No. 20231813.



\noindent{\large Corresponding authors}\\
Correspondence and requests for materials should be addressed to Federico Mazzola, Mario Cuoco, and Wojciech Brzezicki.\\


\noindent{\bf{\large Ethics declarations}}\\
\noindent{\large Competing interests}\\
The authors declare no competing interests.

\noindent{\bf{\large Supplementary Information}}\\
The online version contains Supplementary Information available at https://xxx\\

\end{document}


\title{Supplementary Information:Anomalous spin-optical helical effect in Ti-based kagome metal}

\author{Federico Mazzola**}
\email{federico.mazzola@spin.cnr.it}
\affiliation{CNR-SPIN, c/o Complesso di Monte S. Angelo, IT-80126 Napoli, Italy}

\author{Wojciech Brzezicki**}
\email{w.brzezicki@uj.edu.pl}
\affiliation{Institute of Theoretical Physics, Jagiellonian University, ulic, S. \L{}ojasiewicza 11, PL-30348 Krak\'ow, Poland}
\affiliation{International Research Centre MagTop, Institute of Physics, Polish Academy of Sciences, Aleja Lotnik\'ow 32/46, PL-02668 Warsaw, Poland}

\author{Chiara Bigi}
\affiliation{Synchrotron SOLEIL, L’Orme des Merisiers, D\'epartementale 128, F-91190 Saint-Aubin, France}

\author{Armando Consiglio}
\affiliation{Istituto Officina dei Materiali, Consiglio Nazionale delle Ricerche, Trieste I-34149, Italy}

\author{Luciano Jacopo D'Onofrio}
\affiliation{CNR-SPIN, c/o Universit\'a di Salerno, IT-84084 Fisciano (SA), Italy}

\author{Maria Teresa Mercaldo}
\affiliation{Dipartimento di Fisica ``E. R. Caianiello", Universit\`a di Salerno, IT-84084 Fisciano (SA), Italy}

\author{Adam Kłosiński}
\affiliation{Institute of Theoretical Physics, Faculty of Physics, University of Warsaw, Pasteura 5, PL-02093 Warsaw, Poland}

\author{Fran\c cois Bertran}
\affiliation{Synchrotron SOLEIL, L’Orme des Merisiers, D\'epartementale 128, F-91190 Saint-Aubin, France}

\author{Patrick Le F\`evre}
\affiliation{Univ Rennes, IPR Institut de Physique de Rennes, UMR 6251, F-35000 Rennes, France}

\author{Oliver J. Clark}
\affiliation{School of Physics and Astronomy, Monash University, Clayton, Victoria 3800, Australia}

\author{Mark T. Edmonds}
\affiliation{School of Physics and Astronomy, Monash University, Clayton, Victoria 3800, Australia}

\author{Manuel Tuniz}
\affiliation{Dipartimento di Fisica, Universita degli studi di Trieste, 34127, Trieste, Italy}

\author{Alessandro De Vita}
\affiliation{Fritz Haber Institut der Max Planck Gesellshaft, Faradayweg 4--6, 14195 Berlin, Germany\looseness=-1}

\author{Vincent Polewczyk}
\affiliation{Universit\'e Paris-Saclay, UVSQ, CNRS, GEMaC, 78000, Versailles, France}

\author{Jeppe B. Jacobsen}
\affiliation{Nanoscience Center, Niels Bohr Institute, University of Copenhagen, 2100 Copenhagen, Denmark}

\author{Henrik Jacobsen} 
\affiliation{European Spallation Source ERIC - Data Management and Software Center, 2800 Kgs.~Lyngby, Denmark}

\author{Jill A. Miwa}
\affiliation{Department of Physics and Astronomy, Interdisciplinary Nanoscience Center, Aarhus University, 8000 Aarhus C, Denmark}

\author{Justin W. Wells}
\affiliation{Department of Physics and Centre for Materials Science and Nanotechnology, University of Oslo (UiO), 0318 Oslo, Norway.}

\author{Anupam Jana}
\affiliation{CNR-IOM Istituto Officina dei Materiali, I-34139 Trieste, Italy}

\author{Ivana Vobornik}
\affiliation{CNR-IOM Istituto Officina dei Materiali, I-34139 Trieste, Italy}

\author{Jun Fujii}
\affiliation{CNR-IOM Istituto Officina dei Materiali, I-34139 Trieste, Italy}

\author{Niccolo Mignani}
\affiliation{Dipartimento di Fisica, Politecnico di Milano, Piazza Leonardo Da Vinci 32, Milano 20133, Italy}

\author{Narges Samani Tarakameh}
\affiliation{Dipartimento di Fisica, Politecnico di Milano, Piazza Leonardo Da Vinci 32, Milano 20133, Italy}

\author{Alberto Crepaldi}
\affiliation{Dipartimento di Fisica, Politecnico di Milano, Piazza Leonardo Da Vinci 32, Milano 20133, Italy}

\author{Giorgio Sangiovanni}
\affiliation{Institute for Theoretical Physics and Astrophysics, University of W\"urzburg, D-97074 W\"urzburg, Germany}

\author{Anshu Kataria}
\affiliation{Dipartimento di Scienze Matematiche, Fisiche e Informatiche, Universitá di Parma, I-43124 Parma, Italy}

\author{Tommaso Morresi}
\affiliation{European Centre for Theoretical Studies in Nuclear Physics and Related Areas (ECT*), Fondazione Bruno Kessler, Trento, Italy}

\author{Samuele Sanna}
\affiliation{Department of Physics and Astronomy, University of Bologna, 40127 Bologna, Italy}

\author{Pietro Bonf\'a}
\affiliation{Dipartimento di Scienze Matematiche, Fisiche e Informatiche, Universitá di Parma, I-43124 Parma, Italy}

\author{Brenden R. Ortiz}
\affiliation{Materials Department, University of California Santa Barbara, Santa Barbara, California 93106, USA}

\author{Ganesh Pokharel}
\affiliation{Materials Department, University of California Santa Barbara, Santa Barbara, California 93106, USA}
\affiliation{Perry College of Mathematics, Computing, and Sciences, University of West Georgia, Carrollton, GA 30118, USA}

\author{Stephen D. Wilson}
\affiliation{Materials Department, University of California Santa Barbara, Santa Barbara, California 93106, USA}

\author{Domenico Di Sante}
\affiliation{Department of Physics and Astronomy, University of Bologna, 40127 Bologna, Italy}

\author{Carmine Ortix}
\affiliation{Dipartimento di Fisica ``E. R. Caianiello", Universit\`a di Salerno, IT-84084 Fisciano (SA), Italy}

\author{Mario Cuoco}\email{mario.cuoco@spin.cnr.it}
\affiliation{CNR-SPIN, c/o Universit\'a di Salerno, IT-84084 Fisciano (SA), Italy}

\maketitle

\section{2D Tight-binding model and loop currents states}

\begin{figure}
\includegraphics[width=0.75\columnwidth]{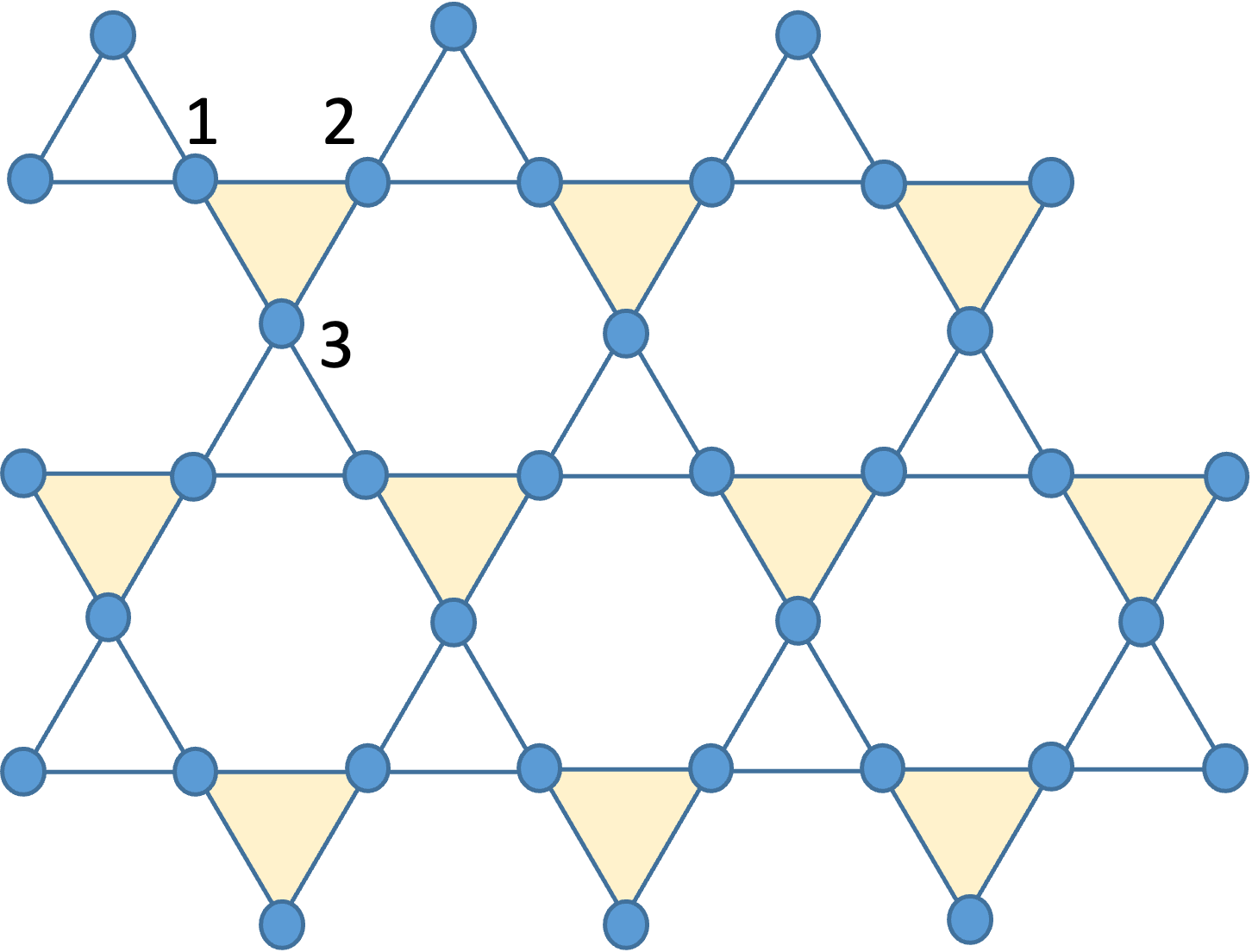}
  \caption{Kagome lattice structure with the labels $\{1,2,3\}$ indicating the positions of the Ti atoms in the unit cell.}
  \label{S0}
\end{figure}

Based on the density functional theory calculations we construct a 2D tight-binding model that is based only on the $d$-orbitals of Ti. To this aim we employ the basis of cubic harmonics $d_{xy}$, $d_{yz}$, $d_{zx}$, $d_{x^{2}-y^{2}}$ and $d_{3z^{2}-r^{2}}$ and thus we need the 
$L=2$ angular momentum matrices to describe the corresponding orbital dependent electronic processes. 
\begin{equation}
    \hat{L}_{x}=\begin{pmatrix}0 & 0 & -i & 0 & 0\\
0 & 0 & 0 & -i & -i\sqrt{3}\\
i & 0 & 0 & 0 & 0\\
0 & i & 0 & 0 & 0\\
0 & i\sqrt{3} & 0 & 0 & 0
\end{pmatrix},\, 
\hat{L}_{y}=\begin{pmatrix}0 & i & 0 & 0 & 0\\
-i & 0 & 0 & 0 & 0\\
0 & 0 & 0 & -i & i\sqrt{3}\\
0 & 0 & i & 0 & 0\\
0 & 0 & -i\sqrt{3} & 0 & 0
\end{pmatrix},\, 
\hat{L}_{z}=\begin{pmatrix}0 & 0 & 0 & 2i & 0\\
0 & 0 & i & 0 & 0\\
0 & -i & 0 & 0 & 0\\
-2i & 0 & 0 & 0 & 0\\
0 & 0 & 0 & 0 & 0
\end{pmatrix}
\end{equation}
Moreover, in order to capture the tri-sublattice structure of the unit cell of the kagome lattice we employ a $T=1$ sublattice pseudospin moment whose components are expressed:
\begin{equation}
\hat{T}_{x}=\begin{pmatrix}0 & 0 & 0\\
0 & 0 & -i\\
0 & i & 0
\end{pmatrix},\quad 
\hat{T}_{y}=\begin{pmatrix}0 & 0 & i\\
0 & 0 & 0\\
-i & 0 & 0
\end{pmatrix},\quad 
\hat{T}_{z}=\begin{pmatrix}0 & -i & 0\\
i & 0 & 0\\
0 & 0 & 0
\end{pmatrix}.
\end{equation}
%
Now, we can define hopping elements between sites $1$ and $2$ of the unit cell as (see Fig. \ref{S0}):
\begin{align}
    \hat{t}_{12}&=\frac{1}{6}dd\sigma\left(\hat{L}_{x}^{2}+\hat{L}_{z}^{2}\right)+\frac{1}{12}\left(5dd\delta+5dd\sigma-8dd\pi\right)\hat{L}_{y}^{2} \nonumber\\
    &+\frac{1}{24}\left(4dd\pi-3dd\sigma-dd\delta\right)\left(\left\{\hat{L}_{x}^{2},\hat{L}_{y}^{2}\right\} +\left\{\hat{L}_{y}^{2},\hat{L}_{z}^{2}\right\} \right)  \nonumber\\  
    &+ia \hat{L}_{z}+ib\left\{ \hat{L}_{y}^{2},\hat{L}_{z}\right\} +ic\left\{ \hat{L}_{x}^{2},\hat{L}_{z}\right\} +d\hat{L}_{z}^{2}+e\left\{ \hat{L}_{y}^{2},\hat{L}_{z}^{2}\right\} +f\left\{ \hat{L}_{z}^{2},\hat{L}_{x}^{2}\right\}. 
\end{align}
Here, $dd\pi$, $dd\sigma$ and $dd\delta$ Slater-Koster parameters are the only ones that are non-vanishing in the case of a fully symmetric simple kagome lattice. However due to the presence of other elements in the lattice (for instance the Bi in the center of the unit cell) also the parameters $a$, $b$, $c$, $d$, $e$, $f$ are non-vanishing, though being small in amplitude.
We also need to define the onsite crystal-field splitting elements for site $3$:
\begin{equation}
        \hat{h}_{3}=q_x\hat{L}_{x}^{2}+q_y\hat{L}_{y}^{2}+q_z\hat{L}_{z}^{2}+r_x\left\{ \hat{L}_{y}^{2},\hat{L}_{z}^{2}\right\}+r_y\left\{ \hat{L}_{x}^{2},\hat{L}_{x}^{2}\right\}+r_z\left\{ \hat{L}_{x}^{2},\hat{L}_{y}^{2}\right\}.
\end{equation}
To get hopping and crystal field terms between other sites in the unit cell it is useful to define a $120\degree$ orbital rotation matrix: $\hat{C}=\exp[i2\pi/3\hat{L}_z]$ (it implicity includes the rotation of the site index within the unit cell). Hence, we have that from the hopping at the bond $12$ and and local terms at one of the site of the unit cell one can construct the Hamiltonian for the other symmetry related bonds and sites:
\begin{equation}
        \hat{t}_{31}=\hat{C}^{\dagger}\hat{t}_{12}\hat{C},\quad \hat{t}_{23}=\hat{C}\hat{t}_{12}\hat{C}^{\dagger},\quad 
        \hat{h}_{2}=\hat{C}^{\dagger}\hat{h}_{3}\hat{C},\quad \hat{h}_{1}=\hat{C}\hat{h}_{3}\hat{C}^{\dagger}.
\end{equation}
Then, we are ready to define a tight-binding Hamiltonian, we have:
\begin{align}
  \hat{H}_{\vec{k}}&=-\left[1+e^{-i\frac{1}{2}\left(k_{x}+\sqrt{3}k_{y}\right)}\right]\hat{t}_{31}\otimes\left(\hat{T}_{x}\hat{T}_{z}\right)-\left[1+e^{i\frac{1}{2}\left(k_{x}+\sqrt{3}k_{y}\right)}\right]\hat{t}_{13}\otimes\left(\hat{T}_{z}\hat{T}_{x}\right) \nonumber\\
   &-\left[1+e^{-ik_{x}}\right]\hat{t}_{21}\otimes\left(\hat{T}_{x}\hat{T}_{y}\right) -\left[1+e^{ik_{x}}\right]\hat{t}_{12}\otimes\left(\hat{T}_{y}\hat{T}_{x}\right)\nonumber\\
   &-\left[1+e^{-i\frac{1}{2}\left(-k_{x}+\sqrt{3}k_{y}\right)}\right]\hat{t}_{32}\otimes\left(\hat{T}_{y}\hat{T}_{z}\right)-\left[1+e^{i\frac{1}{2}\left(-k_{x}+\sqrt{3}k_{y}\right)}\right]\hat{t}_{23}\otimes\left(\hat{T}_{z}\hat{T}_{y}\right)\nonumber\\
   &+\hat{h}_1\otimes\left(1-\hat{T}_{x}^2\right)
   +\hat{h}_2\otimes\left(1-\hat{T}_{y}^2\right)
   +\hat{h}_3\otimes\left(1-\hat{T}_{z}^2\right).
\end{align}
%
The atomic spin-orbit coupling at the Ti site is expressed as $\lambda_{so} \hat{L} \cdot \hat{s}$. For convenience, we have discarded $\lambda_{so}$ since it is relatively small compared to the other energy scales involved in the problem and does not influence significatively the results of the analysis.
The values of the hopping parameters extracted from the Wannier projection analysis are the following (in units of eV):
\begin{align}
    dd\pi&=0.3433,\, dd\sigma=-0.9807, \, dd\delta=0.0267 \nonumber\\
    a&=-0.0894,\, b=-0.0013, \, c=0.0280 \nonumber\\
    d&=0.0759,\, e=-0.0232, \, f=0.0081 \nonumber\\    
    q_x&=0.0720,\, q_y=0.3035, \, q_z=0.0246 \nonumber\\
    r_x&=-0.0124, \, r_y=0.0653, \, r_z=-0.0278. 
\end{align}

To describe the loop current phases, it is advantageous to use current operators that are directly formulated in terms of the sublattice-spin-orbital operators.
The loop current phase with isotropic charge circulation within the unit cell can be expressed by the following symmetry breaking term
\begin{equation}
        \hat{J}_{\rm c}=\left(\hat{T}_x+\hat{T}_y+\hat{T}_z\right) \,,
\end{equation}
since $\hat{T}_x$,$\hat{T}_y$, and $\hat{T}_z$ correspond to the current operators on the three bonds within the unit cell.

Similarly, one can introduce loop current states that possess orbital or spin-orbital quadrupoles. The orbital quadrupoles within the unit cell can be represented by means of the mirror symmetric fields $\hat{A}=(\hat{A}_x,\hat{A}_y,\hat{A}_z)$, with ${\hat{A}}_x=\hat{L}_x \hat{L}_x$, and mirror broken orbital operators $\hat{\xi}=(\hat{\xi}_x,\hat{\xi}_y,\hat{\xi}_z)$, where ${\hat{\xi}}_x=\hat{L}_y \hat{L}_z + \hat{L}_z \hat{L}_y$, with other components derived through index permutation. For the purposes of our work we are interested in the mirror broken loop current state. This configuration can be generally expressed by having symmetry breaking current terms in the Hamiltonian that are superposition of the $\hat{\xi}$ components. Thus, it is expressed as $\hat{J}_o=(\hat{T}_x+\hat{T}_y+\hat{T}_z)(\chi_x \hat{\xi}_x+\chi_y \hat{\xi}_y+\chi_z \hat{\xi}_z)$ with the coefficients $\chi_i$ ($i=x,y,z$) defining the pattern of the orbital quadrupole that breaks rotation and mirror symmetries.
The same approach can be followed when constructing the spin-orbital quadrupoles. In this case one can have different configurations depending on the orientation of the spin and orbital moments. The loop current configuration is consistently established by the charge circulation within the unit cell and is represented through the combiantion of the sublattice operators $\left(\hat{T}_x+\hat{T}_y+\hat{T}_z\right)$. Instead, for the spin-orbital quadrupole, one can have configurations that are either mirror and rotationally symmetric or exhibit broken mirror and rotational symmetry in the spin-orbital space. The symmetric loop current phase is associated to the current operators of the type $\hat{A}_{so}=(\hat{T}_x+\hat{T}_y+\hat{T}_z)(a_x \hat{L}_x \hat{s}_x+a_y \hat{L}_y \hat{s}_y+a_z \hat{L}_z \hat{s}_z)$. These phases do not lead to anomalies in the spin-dichroic and handedness resolved spin-response for the photoemission intensity.
The spin-orbital quadrupole loop current phases that we have considered in the manuscript are marked by cross spin-orbital correlations associated to the components of the term $\hat{\mathbf{L}}\times \hat{\mathbf{s}}$. A representative configuration of this type of spin-orbital quadrupole loop current phase can be then expressed as $\hat{J}_{so}=g_{so} (\hat{T}_x+\hat{T}_y+\hat{T}_z)[\mathbf{n} \cdot (\mathbf{\hat{L}} \times \mathbf{\hat{s}})]$ with $\mathbf{n}=(n_x,n_y,n_z)$ setting the director for the spin-orbital quadrupole distribution.

%
%
%

\section{ARPES and Spin-ARPES data}
ARPES and Spin-ARPES were conducted at both the CASSIOPEE laboratory of SOLEIL (Paris) and the APE-LE laboratory of Elettra (Trieste). The samples were prepared by mounting them in a controlled glovebox environment and top-posting them with ceramic elements affixed using silver epoxy. The silver epoxy was cured within the glovebox at 100$^\circ$C for one hour, after which the samples were transferred into ultra-high vacuum (UHV) conditions, cooled to 15 K, and cleaved \textit{in situ}.

The samples were meticulously aligned along high-symmetry directions, employing various light polarizations and photon energies. Through the analysis of the multi-polarized Fermi surfaces, we achieved azimuthal alignment with an accuracy of 1 degree. The bulk $\Gamma$ point, the focal point of our spin-ARPES investigation, was estimated to occur at approximately 65 eV photon energy. However, as detailed below, all observed bands displayed pronounced two-dimensional characteristics, exhibiting nearly dispersionless features.

\begin{figure}
\includegraphics[width=1.0\columnwidth]{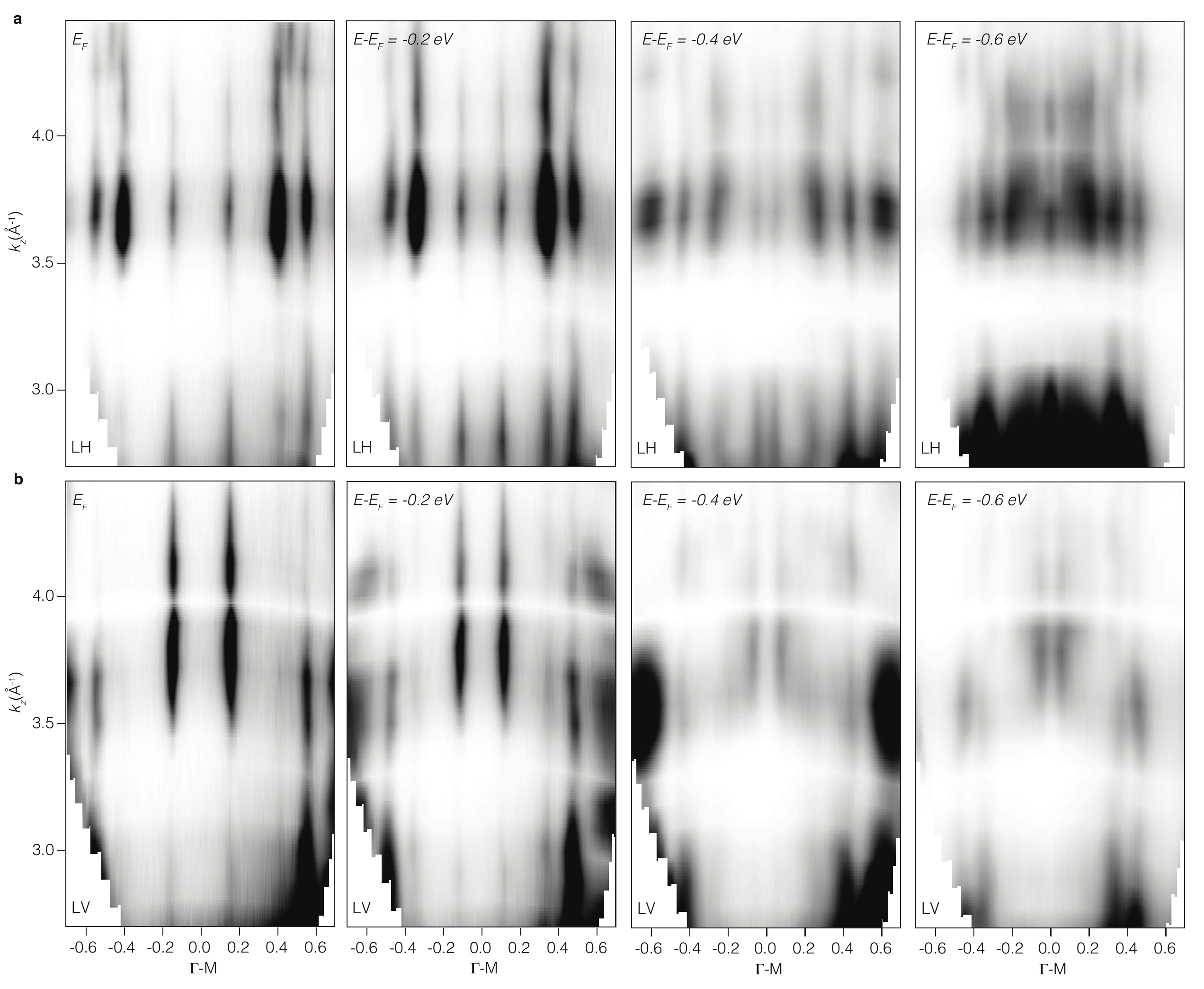}
  \caption{\textbf{a.} Linear horizontal and \textbf{b.} vertical polarizations spectra collected at various photon energies and covering multiple Brillouin zones in $k_z$. The absence of three-dimensionality is confirmed by the stripe-like behaviour shown by the data, which is also independent on the binding energy measured.}
  \label{S1}
\end{figure}

\begin{figure}
\includegraphics[width=0.8\columnwidth]{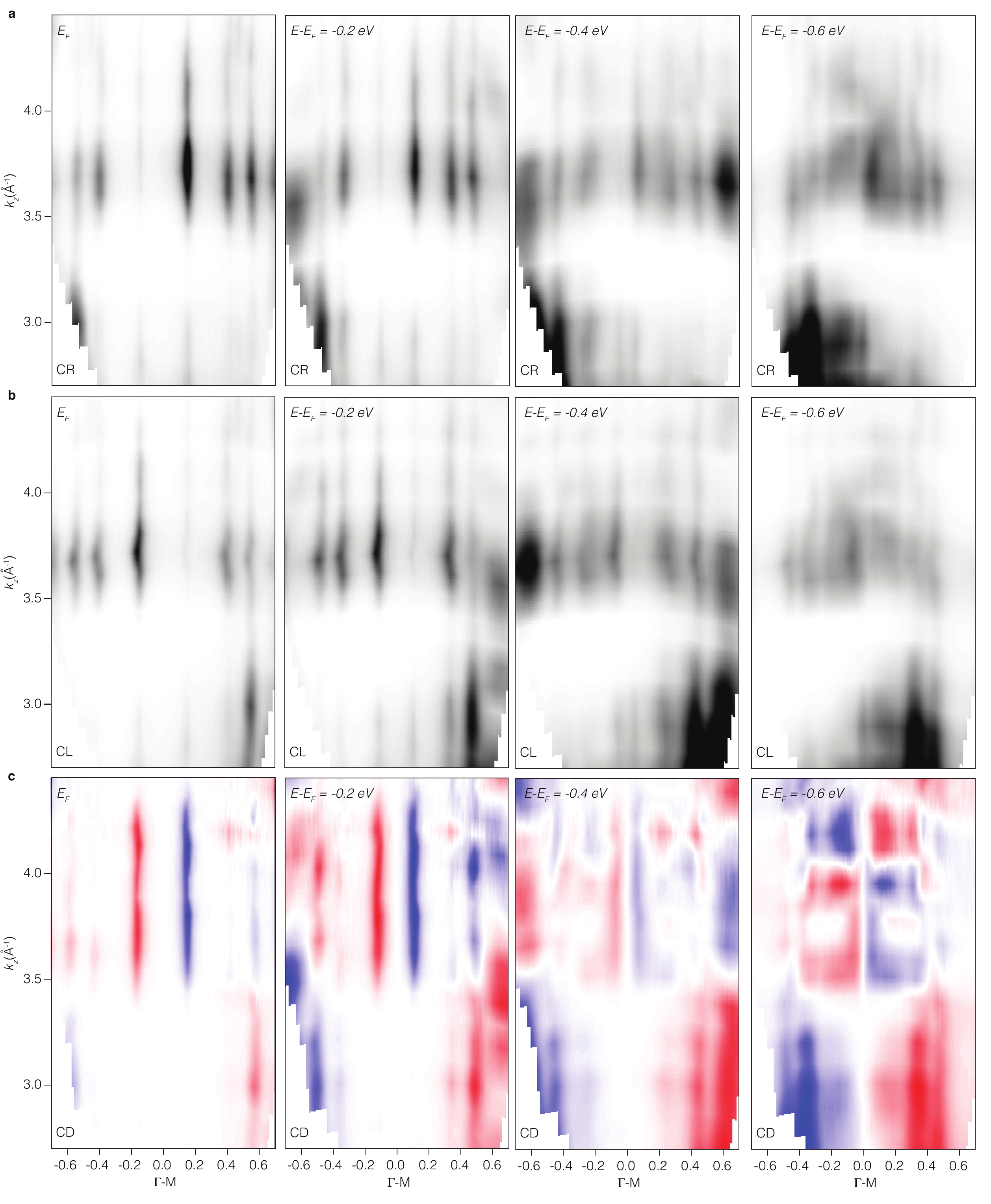}
  \caption{\textbf{a.} Circular right, \textbf{b.} left polarizations, and \textbf{c.}Dichroism spectra were collected across various photon energies, encompassing multiple Brillouin zones in $k_z$. Consistent with the previously presented linear polarizations, the absence of three-dimensional behavior is again observed. Notably, despite the variation in photon energy, the dichroism results are distinctly defined at both positive and negative momenta, with minimal residual contributions at the center of the Brillouin zone. This observation is in alignment with the main text findings and further reinforces the robustness of this effect across different photon energies.}
  \label{S2}
\end{figure}

We commenced our investigation by focusing on spin-integrated ARPES, electronic dimensionality, and polarization control. Spectra were systematically acquired with the analyser slit precisely aligned along both the $\Gamma-M$ direction of the Brillouin zone, across a comprehensive range of photon energies (See Fig. \ref{S1} and Fig. \ref{S2}, for linear and circular polarizations respectively). This methodological approach facilitated a thorough exploration of the system's electronic dimensionality, enabling the identification of any deviations that could indicate three-dimensional dispersion. Notably, the electronic states consistently exhibited a fully two-dimensional character, irrespective of the polarization employed. The same conclusion is drawn by looking at the electronic states below the Fermi level, corroborating the genuine two-dimensional nature of the bands investigated. Furthermore, the extracted circular dichroism at various photon energies exhibits no anomalies along this direction, with only a minimal residual signal observed at the zone center. This finding reinforces the assertion presented in the main text, where the circular dichroism of the spin-integrated signal was shown to be negligible, given the limitations of the experimental resolution.

Another compelling aspect that substantiates the two-dimensional electronic behavior is the lack of significant variations in the dispersion of the system's Fermi surface and constant energy maps. Specifically, while a general redistribution of spectral weight occurs, the fermiology identified across a broad photon energy range exhibits remarkable consistency. This consistency is illustrated in Fig. \ref{S3}-\ref{S4}-\ref{S5}, where no substantial changes are observed across different photon energies. Note that the $\Gamma$ point has been estimated by comparing our data with previous works and by extracting it from the $c$-axis parameters ($c=9.2062$ \AA, and $V_0=6$ eV). Here, for completeness we report Fermi surface maps and constant energy contours collected at various photon energies.

\begin{figure}
\includegraphics[width=0.8\columnwidth]{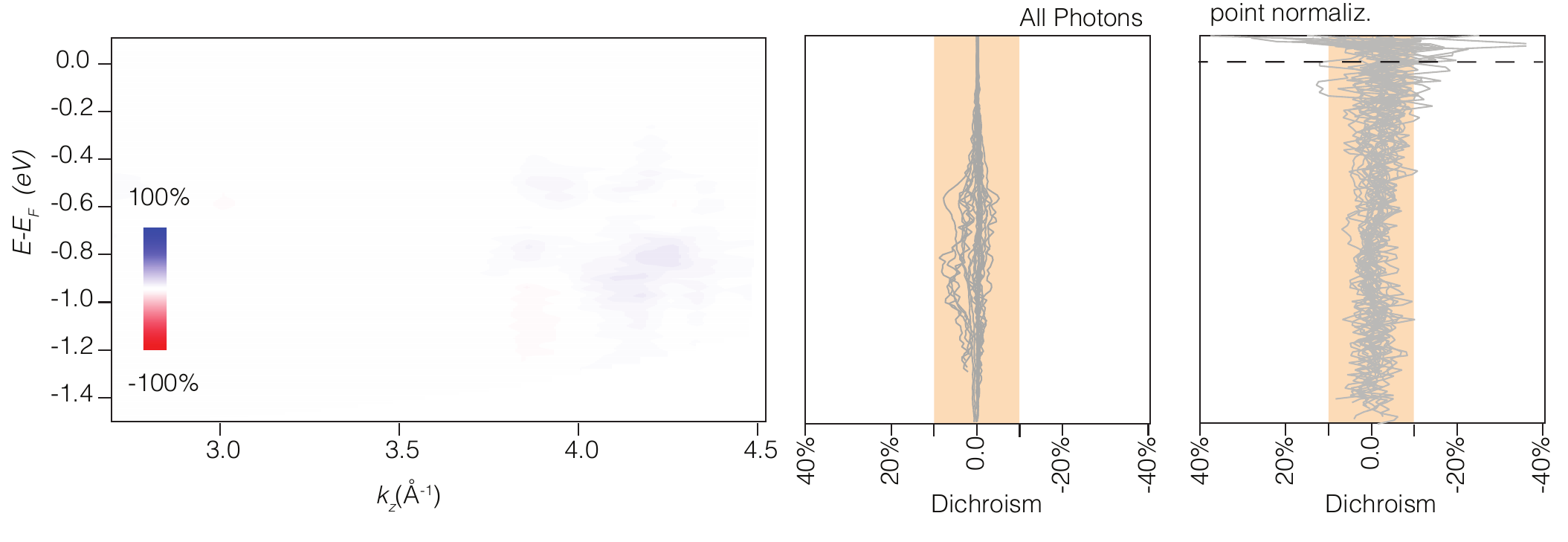}
  \caption{EDCs collected with circular dichroism at several photon energies. As one can see, the maximum deviation occurs with circa 10$\%$ of maximum residual. To the right the single EDCs are reported: the first left line-graph is the standard dichroism obtained as the difference between right and left helicity. The panel to the right shows the EDCs but after the difference in helicities is divided by the sum of them. The increased noise above and in proximity of Fermi is normal and due to a division by a number which is nearly zero. Both methods show clearly the corroboration of a small circular dichroism.}
  \label{Snew}
\end{figure}

\begin{figure}
\includegraphics[width=0.7\columnwidth]{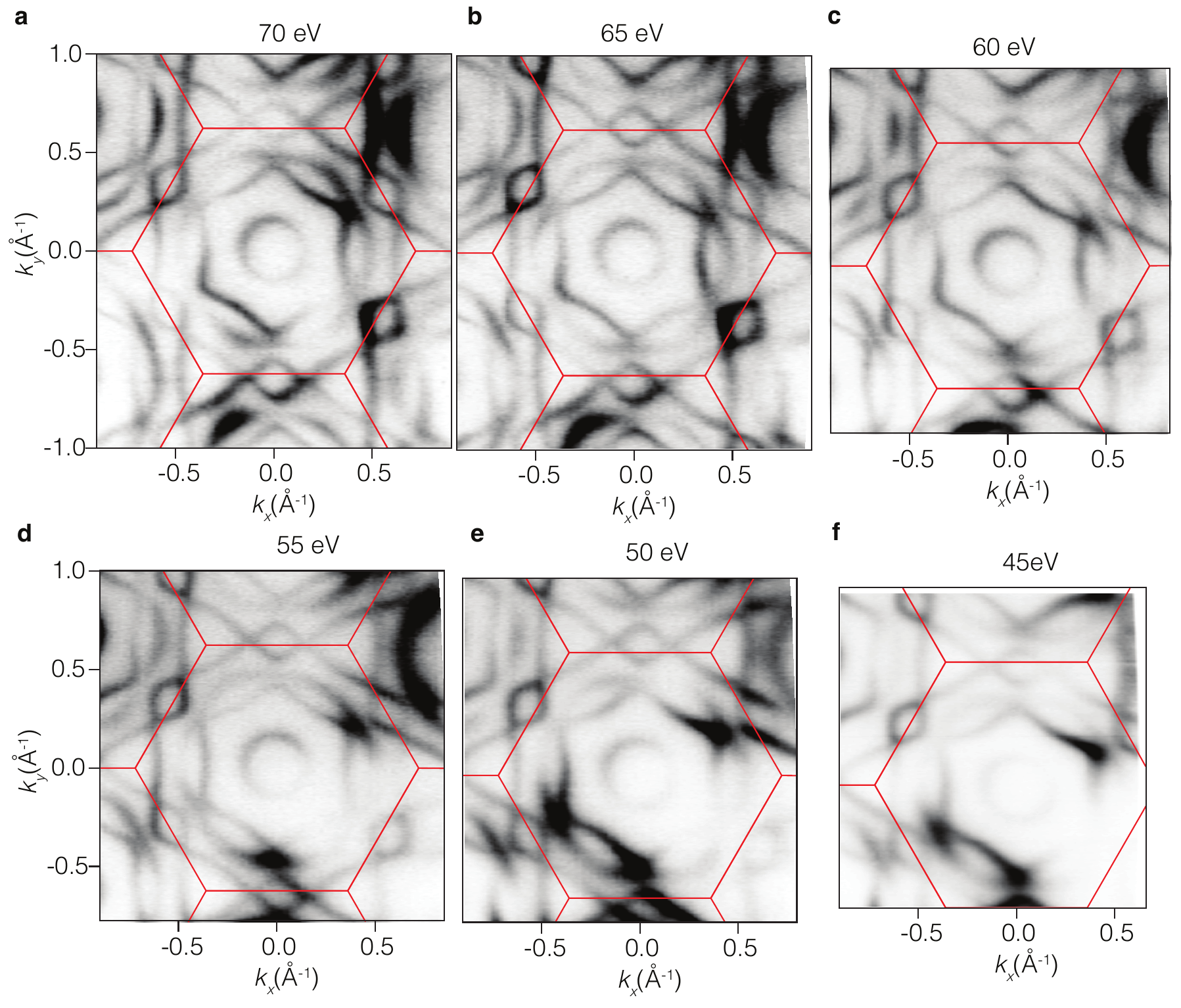}
  \caption{Fermi surface map collected with linear horizontal polarization from 70 eV to 45 eV in step of 5 eV.}
  \label{S3}
\end{figure}

\begin{figure}
\includegraphics[width=0.7\columnwidth]{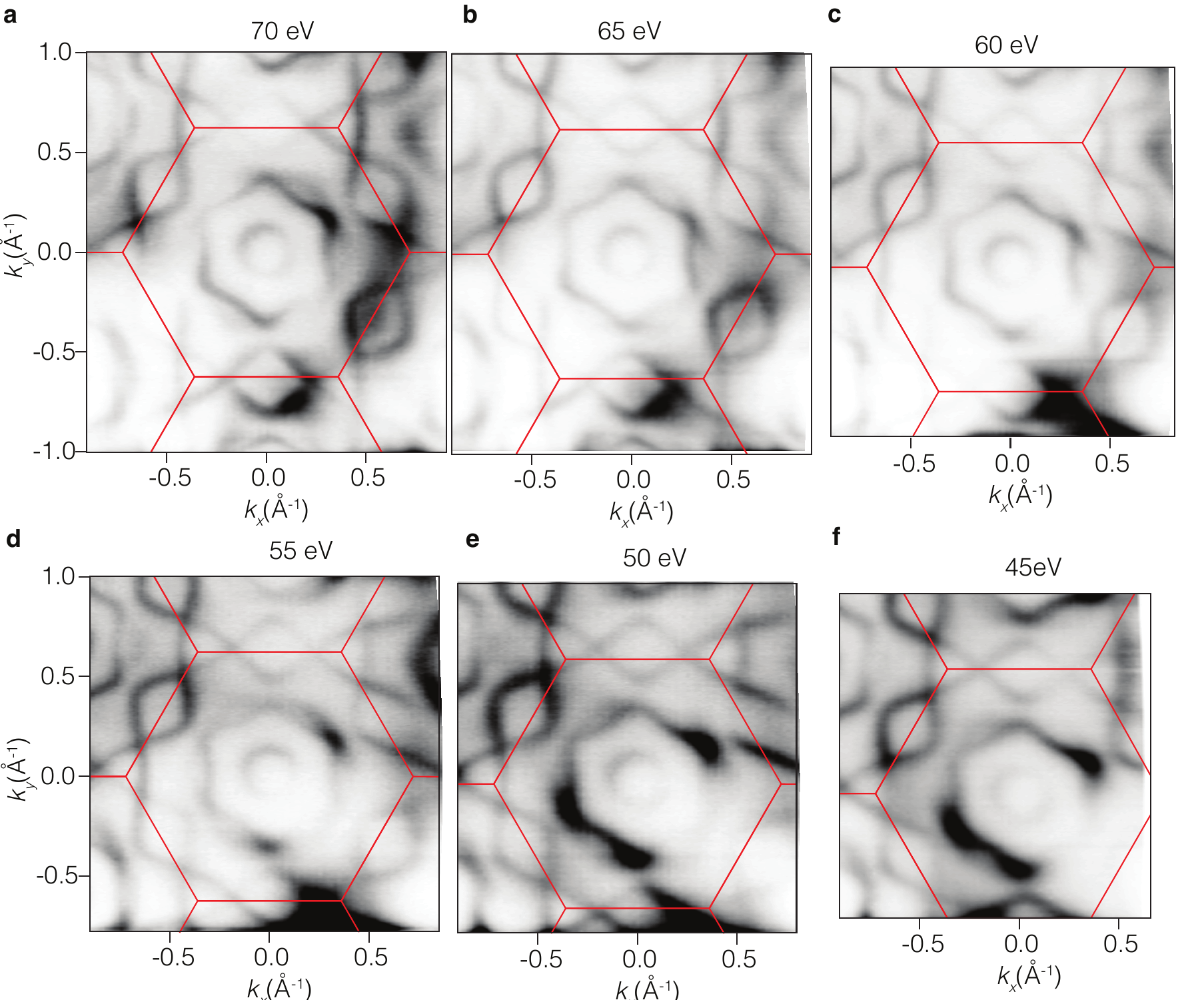}
  \caption{Constant energy contour (200 meV below the Fermi level) collected with linear horizontal polarization from 70 eV to 45 eV in step of 5 eV.}
  \label{S4}
\end{figure}

\begin{figure}
\includegraphics[width=0.7\columnwidth]{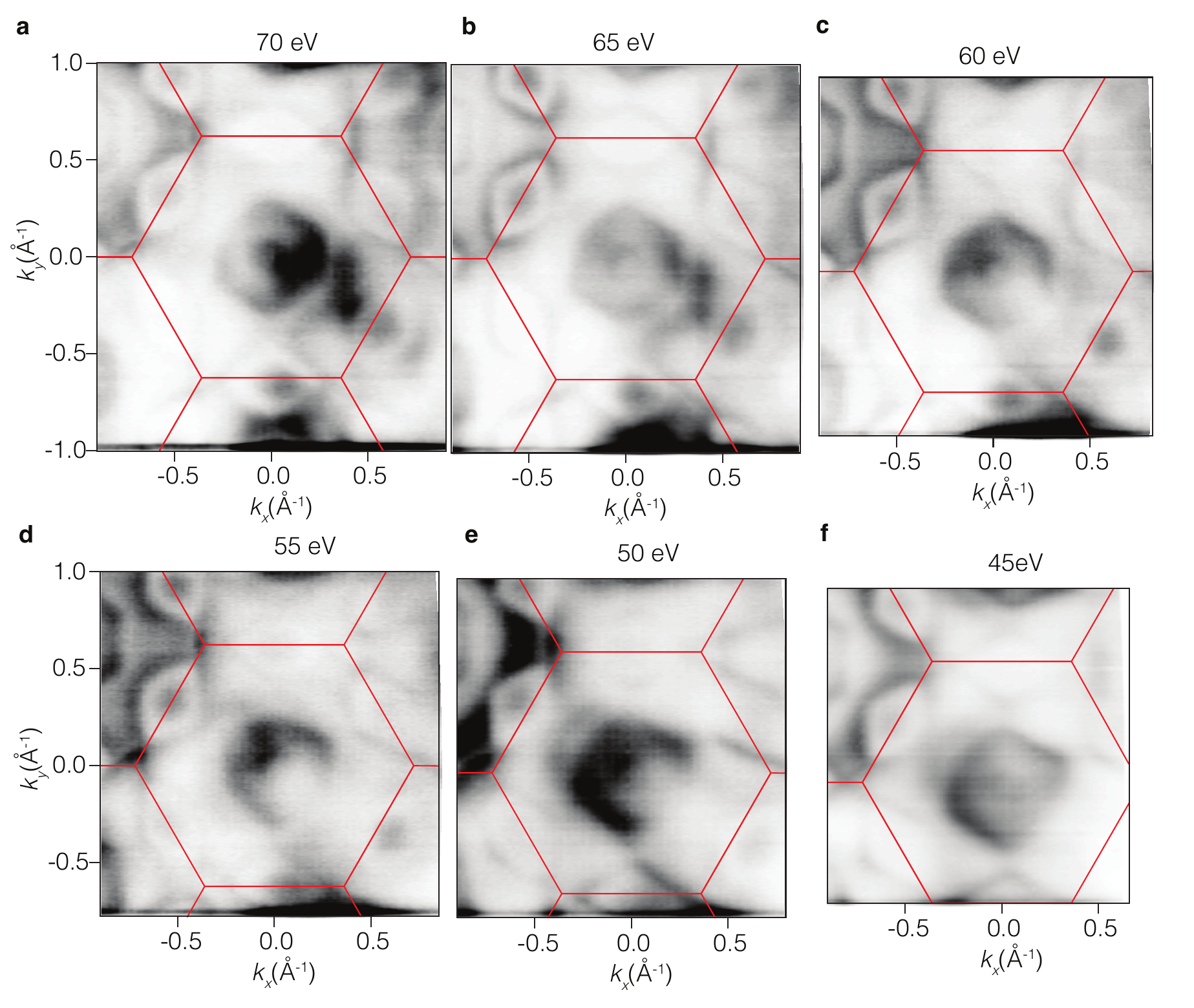}
  \caption{Constant energy contour (400 meV below the Fermi level) collected with linear horizontal polarization from 70 eV to 45 eV in step of 5 eV.}
  \label{S5}
\end{figure}

We now focus on the Spin-ARPES aspect of our study. Initially, using conventional ARPES, we aligned the samples such that the incident light lies within one of the mirror planes of the crystal. In this configuration ($K \to K$ and $M \to M$), the matrix elements are well-defined, enabling us to effectively disentangle the contributions from geometrical matrix elements. Additionally, we concentrated our investigation at the $\Gamma$ point, where the geometrical matrix elements are inherently zero. This strategic approach significantly minimizes the potential for artefacts in our measurements.

\begin{figure}
\includegraphics[width=0.8\columnwidth]{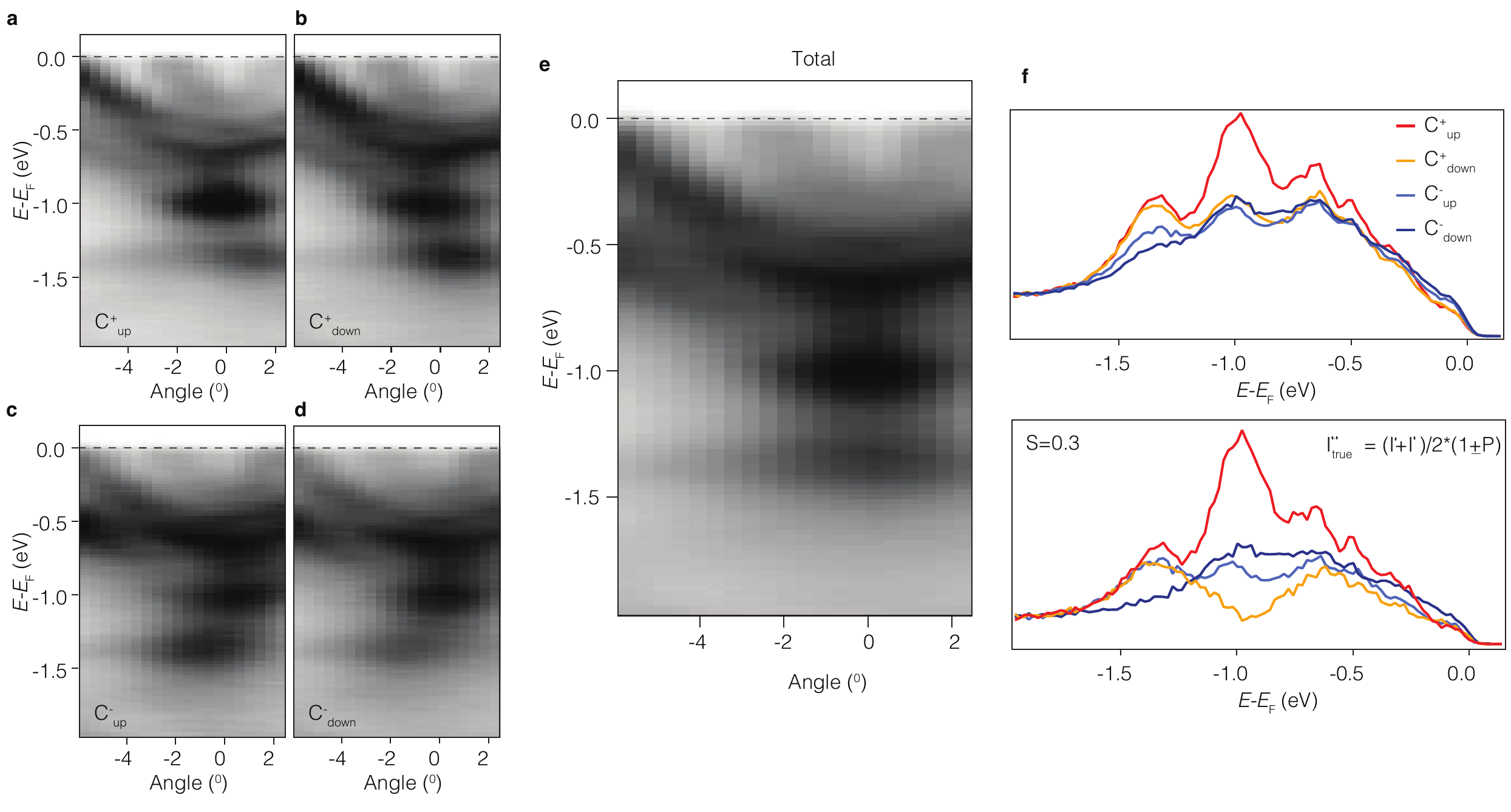}
  \caption{ARPES map reconstructed by using the spin detectors: by collecting EDCs it is possible to build the maps in each polarization and spin and understand the $\Gamma$ point. Data have been collected as \textbf{a} $C^{+}_{\uparrow}$, \textbf{b} $C^{+}_{\downarrow}$, \textbf{c} $C^{-}_{\uparrow}$, and \textbf{d} $C^{-}_{\downarrow}$. Then \textbf{e} the total signal which gives a better overview of the centre of the zone is reconstructed. \textbf{f} EDCs at $\Gamma$ extracted from the raw spectra are shown in the upper panel (note the effect is visible already there) and the ones after Sherman function applied for each helicity and including the spin species at the $\Gamma$ point.}
  \label{S6}
\end{figure}

To acquire spectra at $\Gamma$, we collected photoelectron data at various angles using the spin detector and constructed the ARPES image from these measurements. In Fig. \ref{S6}, we present the spectra obtained solely with the spin detector in various light-polarization configurations, depicted here with a thick grid of points for clarity. The resulting dispersion now reveals the electronic bands as ARPES typically does, allowing us to precisely identify the angle corresponding to the center of the Brillouin zone.

With the $\Gamma$ point identified, we proceeded to acquire spin-resolved energy distribution curves (EDCs) at the zero angle, capturing both spin channels (positive and negative) while employing both right- and left-handed circular polarizations. The spin-ARPES data were normalized by enforcing a uniform background across all polarization and spin channels, ensuring that the signal above the Fermi level was systematically nullified. Following this procedure, we extracted the polarization for each spin species ($\sigma_z=+1$ and $\sigma_z=-1$), incorporating an efficiency factor ($S=0.3$) derived from the Sherman function. Consequently, for each circular polarization, the spin polarization $P$ is given by:

\begin{center}
$P_{C^{+,-}}=\frac{1}{S} \frac{C^{+,-}_{\uparrow} - C^{+,-}_{\downarrow}}{C^{+,-}_{\uparrow} + C^{+,-}_{\downarrow}}$
\end{center}

Upon extraction of the polarization, the EDCs were recalculated to account for the Sherman function, allowing us to express the true spin-resolved intensities as:

\begin{center}
$C^{+,-}_{TRUE}(\uparrow)=\frac{C^{+,-}_{\uparrow} + C^{+,-}_{\downarrow}}{2}*(1+P_{C^{+,-}})$
\end{center}

\begin{center}
$C^{+,-}_{TRUE}(\downarrow)=\frac{C^{+,-}_{\uparrow} + C^{+,-}_{\downarrow}}{2}*(1-P_{C^{+,-}})$
\end{center}

These results correspond to the analysis presented in Fig. 2d of the main text. For the presentation of Fig. 2c, we opted for a representation that provides an intuitive understanding of the degree of spin polarization and dichroic signal. Specifically, the spin-resolved dichroism reflects the percentage of circular polarization for each spin channel, while the spin-integrated dichroism illustrates the residual percentage of the total dichroic signal, where the spin is fully integrated (and thus irrelevant and all summed up). This approach enables for the spin-integrated curves a direct comparison with dichroism data collected via standard ARPES. 

\begin{figure}
\includegraphics[width=\columnwidth]{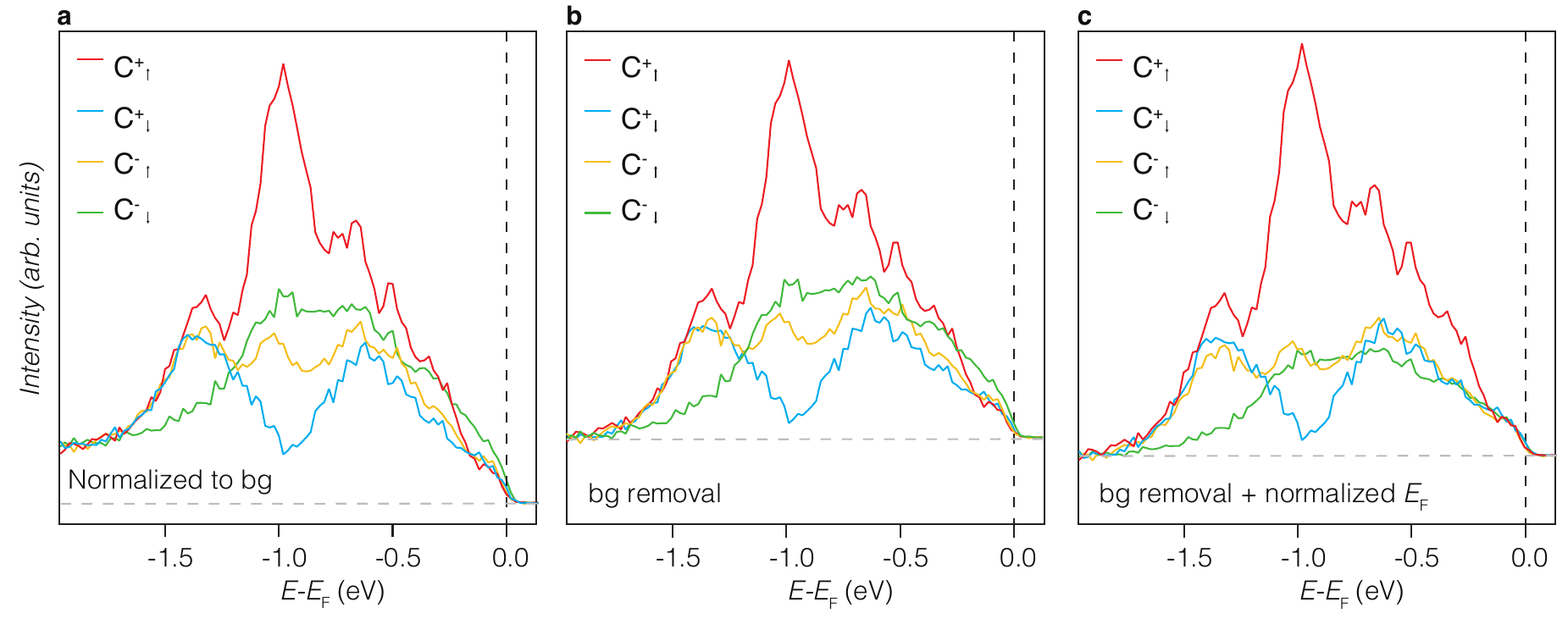}
  \caption{EDCs collected from the $\Gamma$ point after \textbf{a} normalization to the background as in the main text, but here reported on the same graph with the real relative intensity. \textbf{b} The same but after a Shirley background removal and \textbf{c} after forcing the Fermi edge to be the same.}
  \label{S7}
\end{figure}

For completeness, we present the results of alternative analysis approaches in Fig.\ref{S7}. Specifically, we show with normalization the data to their background (as in the main text), then with application of a background subtraction using the Shirley method, and finally, an additional normalization to the Fermi edge.  The most notable observation is that the curves differ significantly, strongly indicating the anomalous spin-optical effect, which remains unaffected by the choice of normalization. For completeness we also extract the polarization of the dichroism and spin-dichroism for Fig.\ref{S7}b-c and we show this in Fig.\ref{S8}: we note that the outcome is entirely independent of the normalization method and procedure applied. The larger oscillations are expected: when dividing by the sum to obtain the percentage ratio, the division by values near zero amplifies the signal. However, the associated uncertainty scales proportionally as well.

\begin{figure}
\includegraphics[width=0.8\columnwidth]{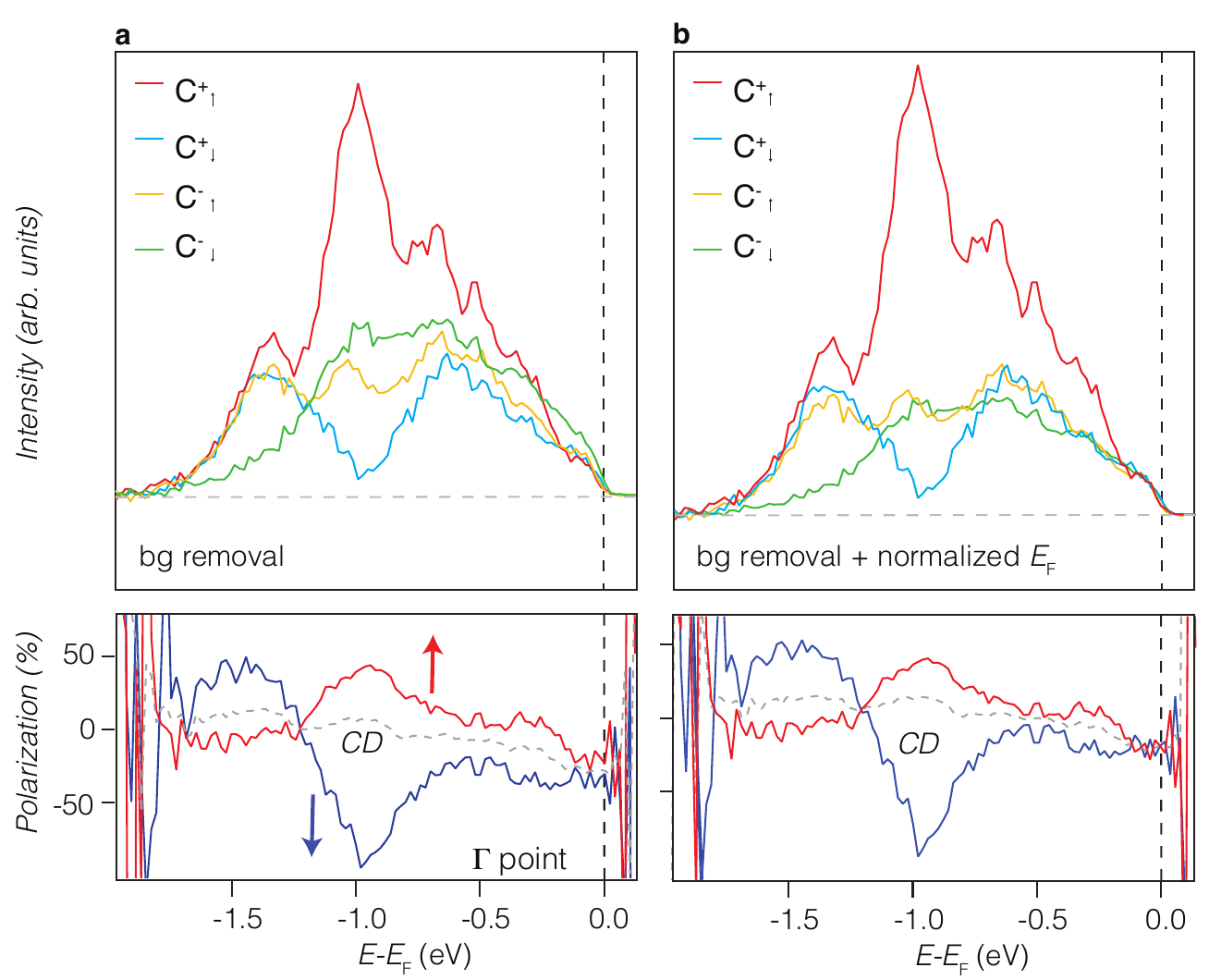}
  \caption{EDCs collected from the $\Gamma$ point after normalization to the background and \textbf{b} after a Shirley background removal and \textbf{c} after forcing the Fermi edge to be the same. Below each panel the extracted spin-dichroic response and canonical circular dichroism are reported. Both methodologies used bring to a similar conclusion. Note that by normalizing around the Fermi level, gives an offset to all the signals.}
  \label{S8}
\end{figure}

\begin{figure}
\includegraphics[width=0.8\columnwidth]{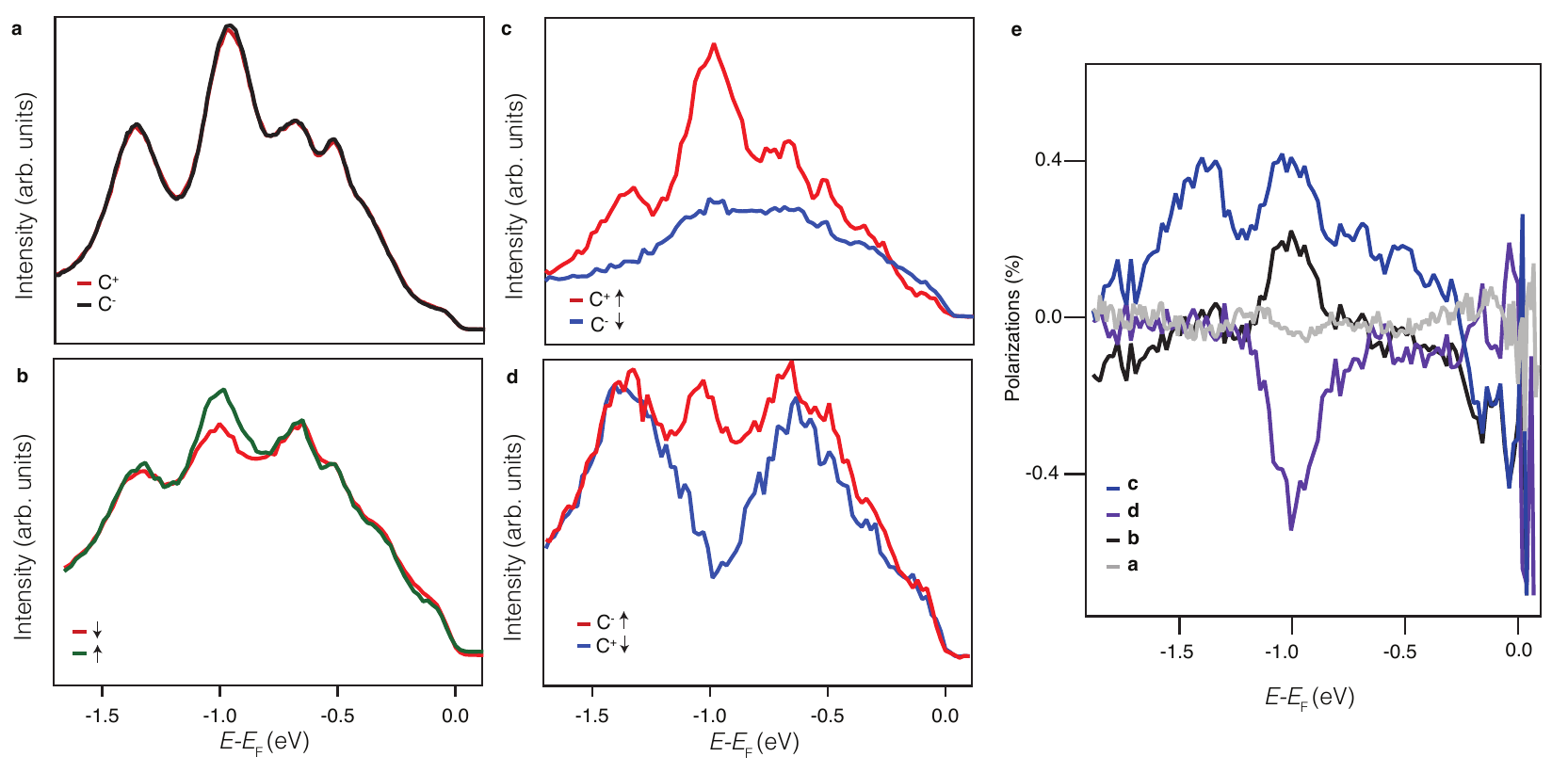}
  \caption{Comparison between spin and dichroism and spin-dichroism and their relative polarization. \textbf{a} Spin-integrated circular positive and negative polarization. The integration performed is of 1 degree and extracted before momentum warping to compare it properly to the spin-resolved data. \textbf{b} Spin-integrated spectra collected with unpolarized light (we summed up circular positive and negative) showing that the spin collected at this point is non-zero but very small and the raw curve show a 5$\%$ difference. \textbf{c} Circular positive and up spin specie and circular negative with species spin down compared and \textbf{d} their mixed compositions. \textbf{e} Percentages in comparison: blue and purple lines show the percentage of spin-dichroic signal meaning the percentage of time-reversal symmetry breaking. The gray curve shows the dichroic difference, which is very small and negligible, while the black is the spin-resolved signal without dichroism. This is more than 4 times smaller and reaches approximately 15$\%$ after dividing it by 0.3 for accounting for a Sherman function. In summary the combined spin and dichroic signal show an amplification.}
  \label{S8}
\end{figure}

Importantly, from a theoretical point of view, the time reversal symmetry breaking should manifest also via analyzing the spin signal alone, without the need of circular polarization, even if its amplitude is expected to be significantly smaller than that one generated by spin-orbital correlations. In Fig.\ref{S8} we show all the various degrees of freedom compared at the $\Gamma$ point. In particular, left and right circular polarization show negligible difference and circularly polarized spin-resolved measurements are instead largely distinct between each others, as shown also in the main text. The spin-resolved measurements (red and green lines - no circular polarization)  show a difference in one of the peaks and negligible difference in the rest of the peaks measured. While this might be compatible with time-reversal symmetry breaking as predicted by theory, we cannot exclude that this state might come from a band aging extremely rapidly within the time-frame of our experiment (we performed all measurements within 30 minutes of total time to avoid this and such a phenomenon was not observed by ARPES within the time frame used and also in reversed ordered for up and down species). Thus, even if precautions were taken, we cannot exclude that the origin of that peak might be an artifact and this would need an investigation by itself. However, the phenomenology observed involves also the other spectral features,  for which spin is negligible or significantly much smaller than the spin-helical signals, corroborating the validity of the match between experiment and theory.

\section{Zero field and Transverse field $\mu$SR results in CsTi$_3$Bi$_5$}

\subsection{Experimental results}

Muon spin rotation and relaxation experiments have been conducted on a powdered sample
mounted on a sample holder covered with a Kapton mask to reduce the background signal.
The experiments were performed using the EMU spectrometer at ISIS, STFC Rutherford Appleton Laboratory, United Kingdom \cite{muons@isis,MUSR}.

\begin{figure*}
    \centering
    \includegraphics[width=1.0\linewidth]{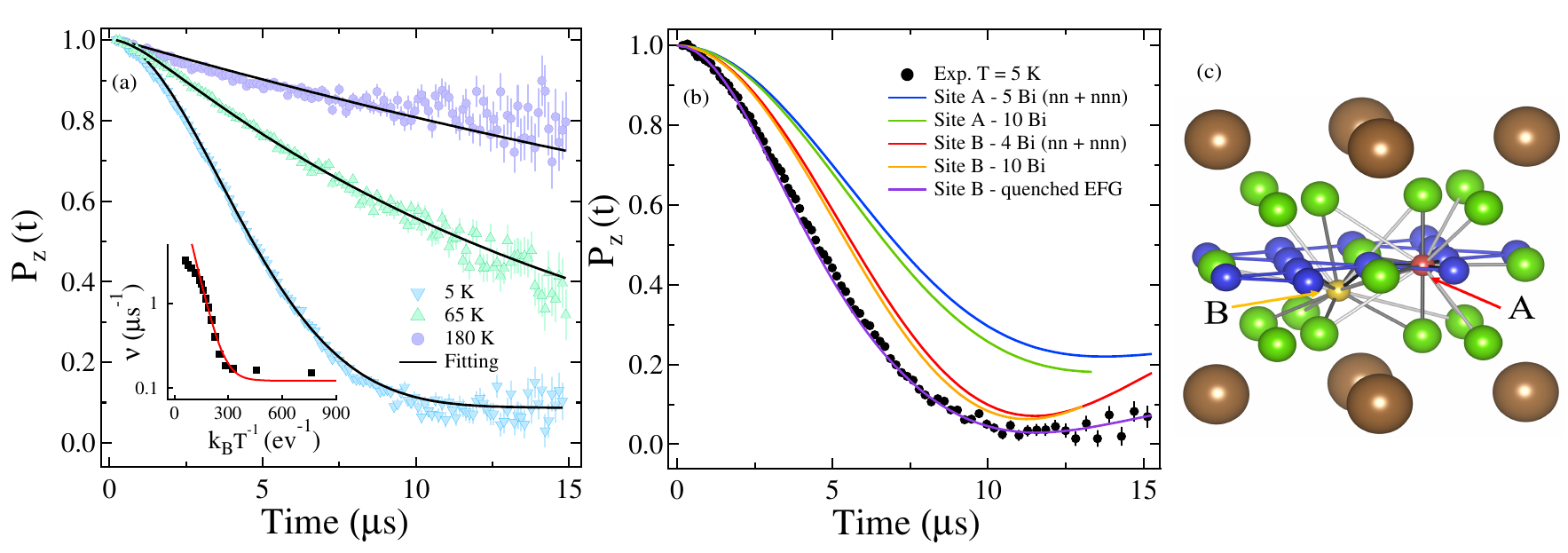}
    \caption{(a) Experimental data at different temperatures are shown, with fitting represented by solid black line. Inset show the Arrhenius law fit for hopping rate $\nu$. (b) Zero-field experimental data at 5~K is compared with the predicted polarisation function for two sites A and B obtained for different sets of nearest neighbors (nn). (c) The two muon sites, site A and B, represented by red and yellow solid spheres, respectively, were obtained using the DFT+$\mu$ method with
    Cs, Ti, and Bi atoms depicted as brown, blue and green spheres, respectively.}
        \label{fig:maintext}
\end{figure*}

The zero field asymmetry spectra obtained at 5~K are shown in Fig.~\ref{fig:maintext}a.
A clear departure from the standard Gaussian Kubo-Toyabe trend is observed.
This can be attributed, in general, to either the anisotropy of the nuclear fields \cite{Solt1995},
or electronic fluctuations or dynamical effects due to quantum tunneling.
Interestingly, with increasing temperature, the ZF-spectrum evolves
toward a marked dynamic behavior, which most likely originates from classical muon diffusion in the lattice. The analysis detailed below shows that the hopping rate $v$ of the muon follows and activated behaviors, as shown in the inset of Fig.~\ref{fig:maintext}a, with an activation energy about 8~meV.

In order to clarify the observed behavior, we have performed first-principles simulations using the Density Functional Theory (DFT)+$\mu$ approach \cite{Blundell_2023, onuorah_2024}. The two lowest energy and almost degenerate stable muon sites are shown in Fig.~\ref{fig:maintext}c while panel b) shows
the predicted relaxation rates obtained after accurate evaluation of perturbation effects introduced by the muon (see SI for details).
Both A and B sites produce a relaxation rate that is slightly slower than the one observed experimentally.
Perfect agreement can be obtained if a large perturbation is induced on the neighboring Bi nuclei, although the expected reduction of the electric field gradient (EFG) is physically unsound.  Further analysis of the quantum effects for the muon in this system is likely required to improve the agreement with the experiment.
Despite the small discrepancy, our results allow us to conclude that no static magnetic fields larger than 0.25~mT can be present at the muon site.
This would potentially correspond to a dipolar contribution from Ti atoms originating from a magnetic moment smaller than $2 \times 10^{-3}$~$\mu_{\text B}$.

\begin{figure}
    \centering
\includegraphics[width=0.45\linewidth]{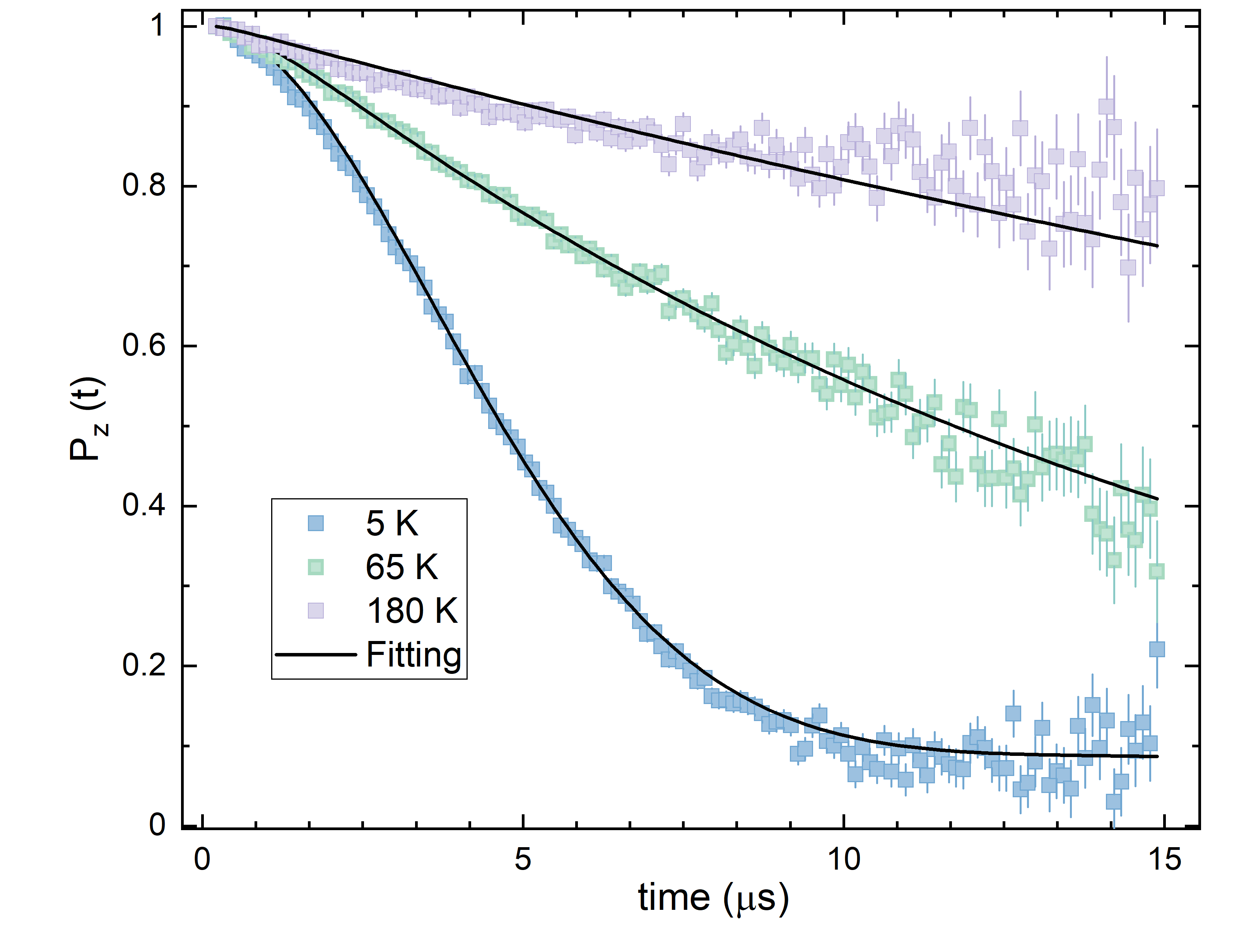}
\includegraphics[width=0.45\linewidth]{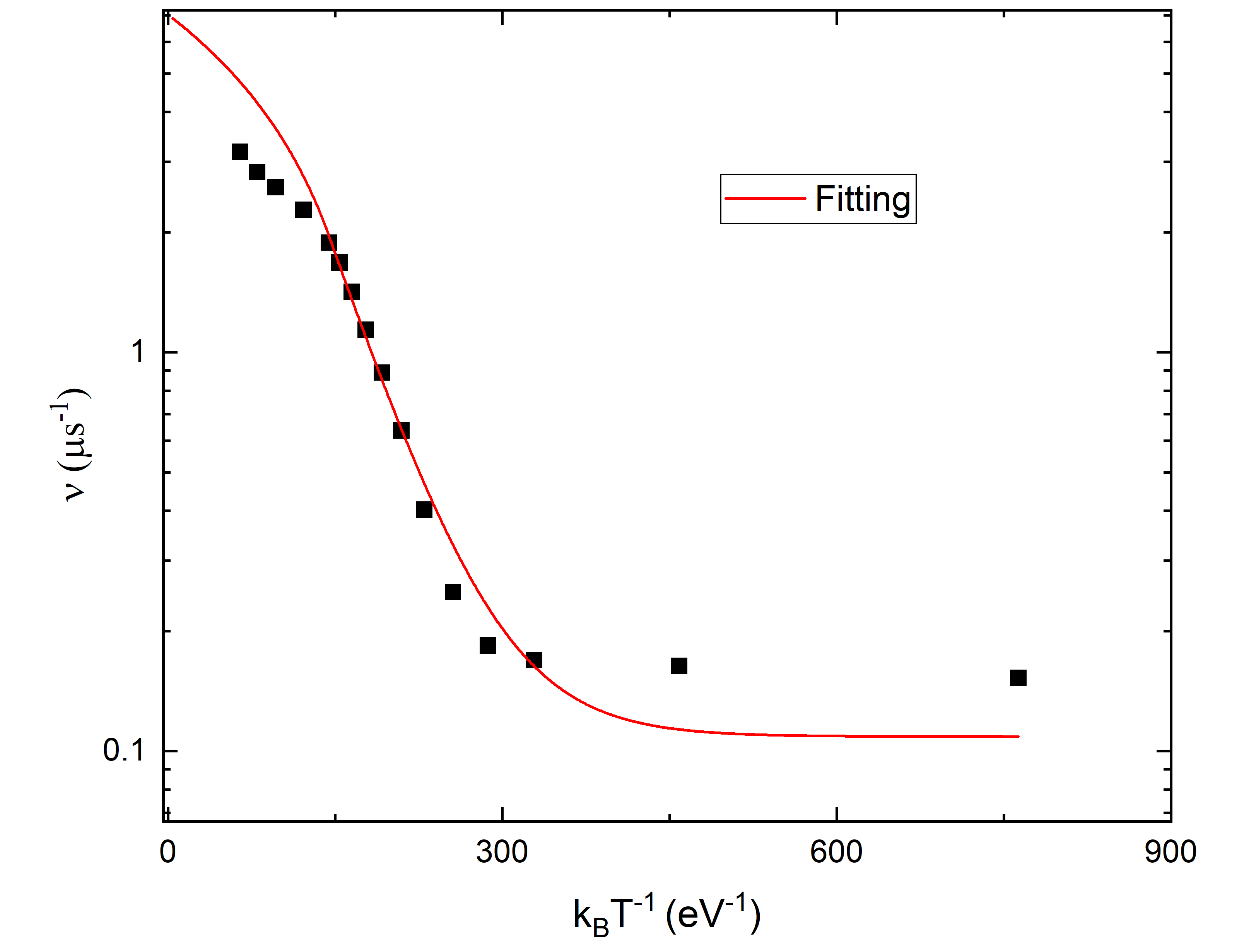}

    \caption{Left: experimental data of CsTi$_3$Bi$_5$ at different temperature where black solid line the corresponding fit of the data using the dynamic Kubo-Toyabe expression. Right:  hopping parameter $\nu$, is plotted against the inverse temperature, along with its fit using eqn. \ref{eq:nu}.}
    \label{fig:exp-nu}
\end{figure}

The zero field muon spin relaxation ($\mu$SR) asymmetry spectra for the lowest measured temperature 5.0~K along with 65~K and 180~K are shown in Fig.~\ref{fig:exp-nu}. The low-temperature data cannot be completely understood by the static Kubo-Toyabe (KT) function. This is due to the approximate nature of this phenomenological fitting function, which does not take into account quantum contributions and anisotropies in the nuclear field distribution \cite{blundell2022muon}.

Moreover, the temperature evolution of spectra depicts the slow depolarization with increasing temperature. To understand the temperature evolution we fit the experimental data using the dynamic Kubo-Toyabe expression, which follows as,
\begin{equation}
    G_z (t) = g_z(t) + \nu \int_0^t g_z(\tau) G_z\left(t-\tau\right) d\tau
\end{equation}
where $g_z(t)$ is the static KT function, and $\nu$ is the muon hopping rate. For zero-field, the static KT function accounts for the randomly distributed Gaussian nuclear magnetic moment and can be written as,
\begin{equation}
g_z(t) = \mbox{A} \Bigg[ \frac{1}{3} + \frac{2}{3} \left( 1 - {\Delta}^2 {t}^2 \right) e^{-\frac{1}{2}\Delta^2 t^2} \Bigg]\label{eq:kt}
\end{equation}
here $\Delta$ represents the second moment of the local field distribution. 
For ZF fitting, the $\Delta$  parameter was fixed to the low-temperature value 0.19498 $\mu$s$^{-1}$, while the hopping rate was set free.
 The temperature dependence of the hopping parameter $\nu$, in terms of its logarithmic value as a function of inverse temperature, is shown in Fig.~\ref{fig:exp-nu}. A thermal activation law is used to evaluate the variation of $\nu$, written as,
\begin{equation}
    \nu(T) = \nu_0 \exp\left(\frac{-E_a}{k_B T}\right)
    \label{eq:nu}
\end{equation}
  where $E_a$ is the activation energy and the estimated value is 8.0(2) meV.  

\subsection{Muon localization}

In order to better understand diffusion mechanisms in the system, we computed the solution to the  Schr\"odinger equation for a muon in the unperturbed lattice. To achieve that, we discretized the muon Hamiltonian employing a finite-difference method with periodic boundary conditions. 
By using an orthorhombic supercell with lattice parameters $a=11.739$ \text{\AA}, $b=10.166$ \text{\AA} and $c=18.556$ \text{\AA}, the Potential Energy Surface (PES), i.e., the potential felt by the muon inside the crystal, has been sampled by placing the muon in a regular grid of $31 \times 31 \times 39$ points along the $x,y$ and $z$ directions respectively. Due to the symmetries of the crystal, this resulted in 830 irreducible (or independent) points that we evaluated using DFT as implemented within Quantum Espresso~\cite{qe_2017} (version 7.1). The 830 different DFT calculations have been performed using a wavefunction cutoff of 70 Ry, a smearing parameter equal to 0.01 Ry and a $4 \times 4 \times 3$ k-points grid. By relaxing the cell, we found that the best agreement between experimental data~\cite{Chen_2023,Yang_2024} and simulations is obtained using a rVV10 non-local correlation functional~\cite{Sabatini_2013}.
The lowest eigenvalues and corresponding eigenvectors of the discretized 3-dimensional single-particle Hamiltonian have then been computed using the Arnoldi package \cite{arpack} for large sparse matrices. The mass of the particle is $m_{\mu}=206.768$ atomic units.   

\begin{figure}
    \centering
    \begin{tikzpicture}
    \draw (0, 0) node[inner sep=0] {\includegraphics[width=8cm]{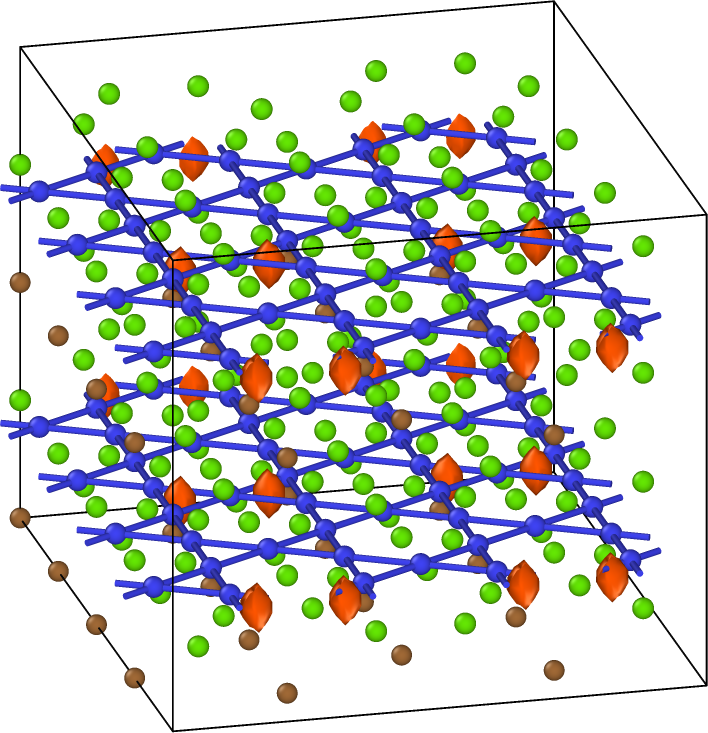}};
    \draw (-4, 4) node {a)};
\end{tikzpicture}
    \begin{tikzpicture}
    \draw (0, 0) node[inner sep=0] {\includegraphics[width=8cm]{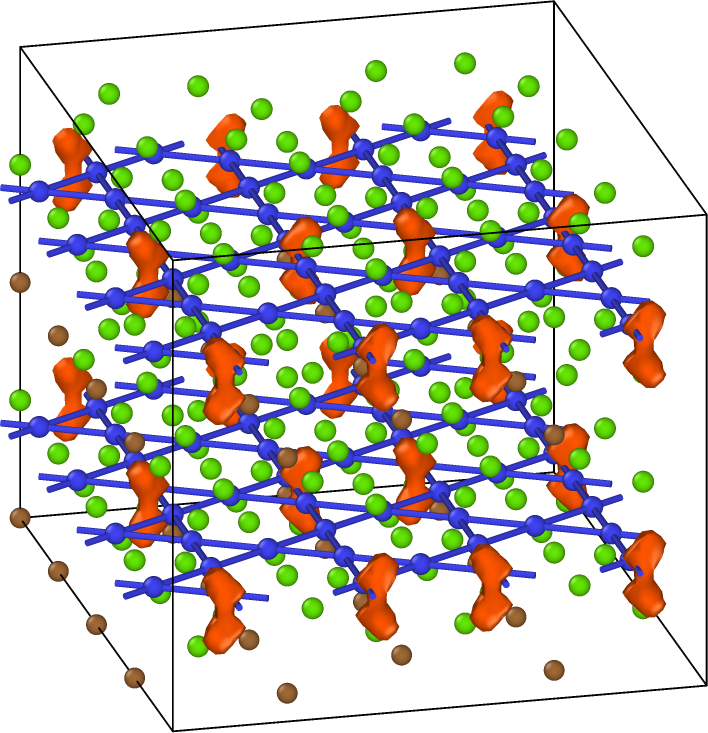}};
    \draw (-4, 4) node {b)};
\end{tikzpicture}
    \begin{tikzpicture}
    \draw (0, 0) node[inner sep=0] {\includegraphics[width=8cm]{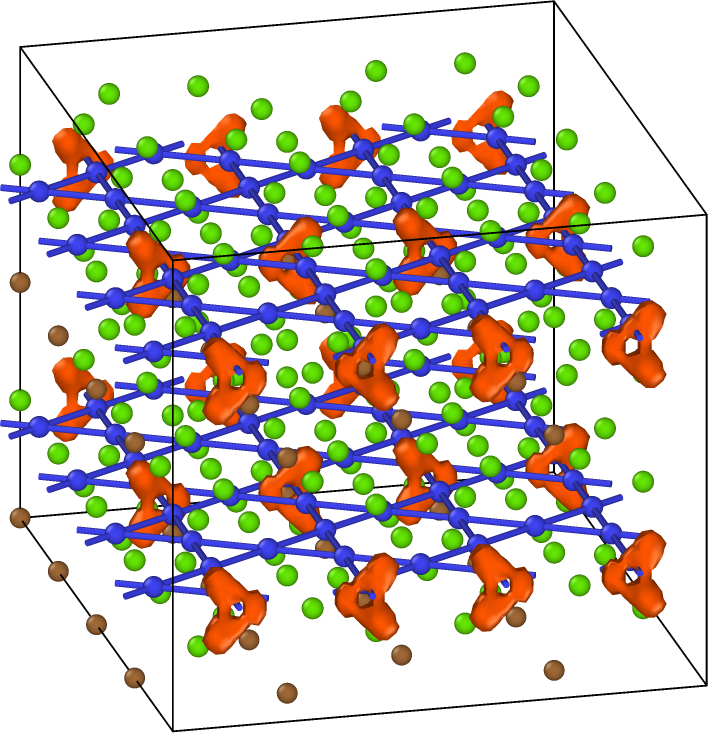}};
    \draw (-4, 4) node {c)};
\end{tikzpicture}
    \caption{Square modulus of the muon wavefunction in \ctb{} (with Ti as blue spheres, Cs brown and Bi green) for the lowest energy eigenstates. Only few states for each set of degenerate solutions (corresponding to equivalent positions in the lattice) are shown. The isosurface shows the value matching 1 \% of the maximum probability density. \label{fig:site-sch}}
\end{figure}

As it can be seen from Fig.~\ref{fig:site-sch}a, the ground state corresponds to a muon wavefunction well localized between three Ti atoms and it is exactly aligned with the kagome plane along the $c$ direction.
The first excited state is instead slightly below and above the Ti plane as shown by Fig.~\ref{fig:site-sch}b.
These results show that a localized ground state and a slightly more delocalized excited state exist and have non-overlapping wavefunctions. The two states are separated in energy by only 140~meV and are therefore both likely to be populated.
A third eigenstate, shown in panel c) and 148~meV above the ground state, has instead a probability density that extends over both the previous sites.
These results show indeed that a number of almost degenerate state exist in the kagome plane and can easily lead to muon diffusion as the temperature raises.

\subsection{First principles description of muon sites}

\begin{figure}
    \centering
\includegraphics[width=0.5\linewidth]{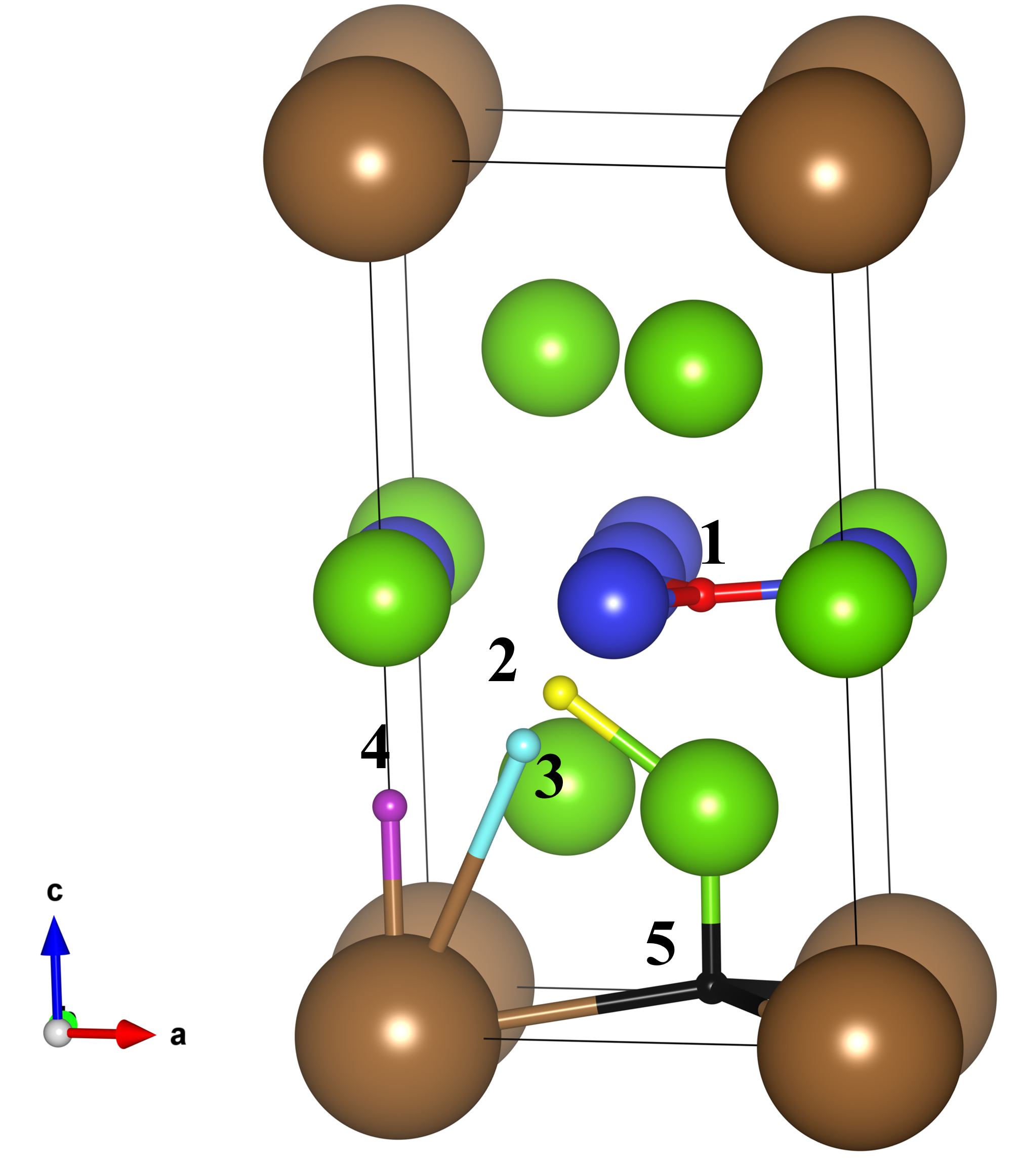}
    \caption{Muon sites in CsTi$_3$Bi$_5$}
    \label{fig:sites}
\end{figure}

The refined position of muon sites in this system and the perturbation produced by the positive interstitial is computed using the approach described by Lancaster and Blundell \cite{Blundell_2023} with the method and the code described in Ref.~\cite{onuorah_2024}. 
A 2x2x1 cell was initially used while refined simulations with a 4x4x2 supercell are performed for the two lowest energy sites.
The results obtained after the structural relaxation are shown in table~\ref{tab:eng_tab} and visualized in the unperturbed unit cell in Fig.~\ref{fig:sites}.
Notably, the two lowest energy sites correspond to the lowest energy eigenstates obtained from the solution of the 3D Schr\"odinger equation, indicating that self-trapping effects are rather limited in this system.

For each site, the perturbation induced by the muon on both the lattice structure and the electric field gradient (EFG) of neighboring atoms is computed and used to produce the polarization functions discussed in the next section.
We note that the EFG at Bi sites is very large (see table~\ref{tab:efg}) and we therefore expect the Bi nuclei to be in the large quadrupole splitting regime.
We also report that the potential along the $z \parallel c$ for site 1 is anharmonic but symmetric with respect to the kagome plane. This can also be appreciated in the results of Fig.~\ref{fig:site-sch}. As a consequence, the expectation value for the position operator along the $z$ is not altered by the anharmonicity of the potential nor it displays relevant temperature dependencies.

\begin{table}[]
    \centering
    \begin{tabular}{|c|c|c|c|c|}
    \hline
     Label & Position (frac. coord) & Energy difference (eV) & Distance from nn ({\AA}) & Distance from nnn ({\AA})\\
     \hline
       1 - A &(0.332,0.667,0.5)& 0  &  1.7538 (Ti)& 2.5676 (Bi i)\\
       2 - B  & (0.372,0.184,0.384) &0.108  & 1.8961 (Ti)&  3.3895 (Bi o)\\
       3 & (0.277,0, 0.342)& 0.558 &   1.8649(Ti)&  2.4984 (Bi o)\\ 
      4 & (0,0,0.270)&0.634 & 1.8454 (Bi o)& 2.7068 (Cs)\\
      5 & (0.333,0.667,0.023)&1.034 & 2.0039 (Bi o)& 3.4131 (Cs)\\
       \hline
    \end{tabular}
    \caption{Total energy of the candidate muon sites found in \ctb{} along with its distance from nearest neighbor (nn) and next nearest neighbor (nnn). The two lowest energy sites are reported with letters A and B in the main text. The fraction coordinates and Wyckoff Symbol of other atoms are as Cs:(0,0,0; 1a),  Ti (0.5,0.5,0.5;3g) and Bi i (0,0,0.5; 1b) and Bi o (0.333,0.667,0.249;4h).   }
    \label{tab:eng_tab}
\end{table}

\subsubsection{Polarization function}

For each site, the muon polarization function is obtained by computing the time evolution of the muon spin according to the following Hamiltonian

\begin{equation}
    \mathcal{H} = \mathcal{H}_{dip,i} + \mathcal{H}_{Q,i} 
\end{equation}
where
\begin{eqnarray}
\mathcal{H}_{dip,i} &=& \frac{\mu_0 \hbar^2 }{4\pi}\gamma_{i}\gamma_{\mu} \left( \frac{\mathbf{I}_{i}\cdot\mathbf{I}_{\mu}}{r^3} -  \frac{3(\mathbf{I}_{i}\cdot\mathbf{r})(\mathbf{I}_{\mu}\cdot\mathbf{r})}{r^5} \right)\\
\mathcal{H}_{Q,i} &=& \frac{eQ_{i}}{6I_{i}(2I_{i}-1)} \sum_{{\alpha},{\beta}{\in}\{{x,y,z}\}} V_{i}^{{\alpha}{\beta}} \left[ \frac{3}{2}\left( I_{i}^{\alpha}I_{i}^{\beta} - I_{i}^{\beta} I_{i}^{\alpha}  \right) - \delta_{{\alpha}{\beta}}I_{i}^{2} \right]
\end{eqnarray}

$\mathcal{H}_{dip,i}$ is the dipolar interaction between the muon and the nuclei accounting for the perturbation induced by the muon on the lattice site and $\mathcal{H}_{Q,i}$ is the quadrupolar interaction between muon and nuclear spin, with $V_i$ being the EFG at nuclear site $i$. The remaining factors and constants should be clear from the context.
The electric field gradient at the various nuclear sites is calculated with PAW pseudopotentials with the GIPAW code \cite{gipaw}.
The values reported in Tab.~\ref{tab:efg} for the unperturbed structure are only slightly affected by the presence of the muon, generally by less than a factor 2. 

The numerical solution to the time evolution of the muon spin is obtained with the approach proposed by Celio \cite{Celio_1986} as implemented in the UNDI code~\cite{undi}. 

\begin{table}
    \centering
    \begin{tabular}{c|c|c}
         &  $|V_{zz}|$& $\eta$\\
         \hline
         Bi1&  6.73& 0\\
 Bi2& 6.22&0\\
         Ti&  0.56& 0.13\\
 Cs& 0.17&0\\
    \end{tabular}
    \caption{Unperturbed EFG in CsTi$_3$Bi$_5$. Values for $V_{zz}$ are in atomic units, the conversion factor to SI is $9.717 \times 10^{21}$~V/m$^2$.}
    \label{tab:efg}
\end{table}
The convergence of the polarization function is limited by the large spin of Bi nuclei ($I=9/2$).
In our analysis we exclude Ti atoms, since only $\sim$15\% of Ti nuclei have a (small) nuclear moment.
Ignoring titanium, the set of nearest (nn) and next nearest neighbor (nnn) atoms includes 5 Bi for site 1 (or A) and 4 Bi for site 2 (B).
A simple strategy to estimate how far is convergence for a given set of nuclei included in the simulation is using the second moment of the nuclear dipole field distribution, given by
\begin{equation}
\sigma_{\infty}^{2} = \frac{2}{3} ( \frac{\mu_{0}}{4 \pi} )^{2} \hbar^{2} \gamma_{\mu}^{2}  \sum _{ j = 1 } ^N  \frac{\gamma_{j}^{2} I_{j} ( I_{j} + 1 )}{ r_{j} ^{6}} =  \sigma_{incl}^{2} + \sigma_{excl}^{2} \label{eq:sigma}
\end{equation}
where $N$ is the total number of nuclei (including Ti and Cs isotopes) in the sample and we separate the contribution included in the numerical simulation, $\sigma_{incl}$ from the total.

The square root of the ratio between the terms included in the simulation ( $\sigma_{incl}$) and $\sigma_{\infty}$ is 0.92 for site A and 0.94 for site B, thus showing that, despite the small set of neighbors included in the calculation, the result is less than 10\% away from convergence.
To further check this point we computed for sites 1 and 2 (A and B in the main text) the evolution of the muon spin considering 10 neighboring Bi atoms of the muon. The result is slightly improved, but is still not aligned with the experimental trend.

\begin{figure}
    \centering
    \includegraphics[]{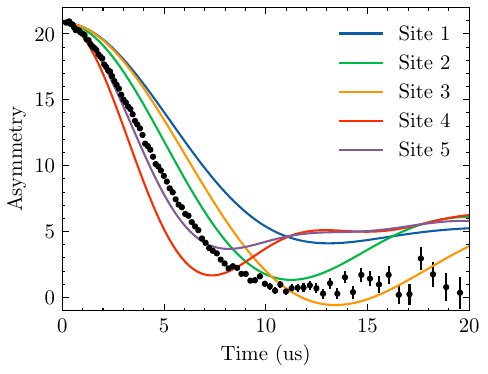}
    \caption{Calculated polarization function obtained from first principles simulations.}
    \label{fig:pols}
\end{figure}

Surprisingly, none of the five interstitial sites shows good agreement with the experiment, as shown in Fig.~\ref{fig:pols}.
A very good agreement is obtained instead when site 2 is considered with a quenched EFG on Bi atoms, as shown in the same figure. The predicted polarization function works well at both short and long time, but the EFG value set on Bi atoms to produce this curve is 3 orders of magnitude smaller than the value predicted by density functional theory simulations.
While such a small value is unrealistic, our results suggest a non-trivial interaction between the muon and the hosting electronic system whose description is beyond the scope of the current study.

The discrepancy between the predicted relaxation rate and the experimentally observed one is about 0.06 $\mu$s$^{-1}$ and the missing contribution can be obtained with local magnetic fields at the muon site smaller than 0.2~mT.
Even in the non-physical assumption that the relaxation rate $\Delta$ of the Kubo-Toyabe function (Eq.~\ref{eq:kt}) is entirely of electronic origin, the magnetic field at the muon can be estimated from the maximum of the Maxwellian field distribution for  producing the experimental relaxation rate. The result turns out to be 0.3~mT, which corresponds to local magnetic moments at the Ti sites smaller than one hundredth of Bohr magneton.

\bibliography{kagome}